\documentclass[12pt]{article}

\usepackage{verbatim}
\usepackage{amsmath}
\usepackage{amssymb}
\usepackage{graphicx}
\usepackage{cite}
\usepackage{subfig}
\usepackage{setspace}
\usepackage{url}
\usepackage[percent]{overpic}
\usepackage{slashed}
\usepackage{authblk}

\usepackage{xspace}
\newcommand{\unit}[1]{\ensuremath{\mathrm{\,#1}}\xspace}

\newcommand{\infb}{\unit{fb^{-1}}}

\usepackage{fullpage}
\usepackage{hyperref}
\def\tev{\,{\rm TeV}}
\def\gev{\,{\rm GeV}}
\def\ie{{\it i.e.}}
\def\eg{{\it e.g.}}

\def\etal{{\it et al.}}

\def\to{\rightarrow}

\title{Higgs Coupling Measurements and Direct Searches as Complementary Probes of the pMSSM}
\date{July 25, 2014}
\author[a]{M. Cahill-Rowley}
\author[a]{J. Hewett}
\author[b,c]{A. Ismail}
\author[a]{T. Rizzo}
\affil[a]{SLAC National Accelerator Laboratory, Menlo Park, CA, USA\footnote{mrowley,hewett,rizzo@slac.stanford.edu}}
\affil[b]{Argonne National Laboratory, Argonne, IL, USA\footnote{aismail@anl.gov}}
\affil[c]{ and University of Illinois at Chicago, Chicago, IL, USA}

\begin{document}

\rightline{\vbox{\halign{&#\hfil\cr
&SLAC-PUB-15935\cr
%&DRAFTv3.0\cr
}}}

%\maketitle

{\let\newpage\relax\maketitle}

\begin{abstract}
The parameter space of the MSSM can be probed via many avenues, such as by precision measurements of the couplings of the $\sim 126$ GeV Higgs boson, as well as
the direct searches for SUSY partners.  We examine the connection between these two collider observables at the LHC and ILC in the 
19/20-parameter p(henomenological)MSSM. Within this scenario, we address two questions: ($i$) How will potentially null direct searches for SUSY at the
LHC influence the predicted properties of the lightest SUSY Higgs boson? ($ii$) What can be learned about the properties of the superpartners from precision measurements
of the Higgs boson couplings? In this paper, we examine these questions by employing 
three different large sets of pMSSM models with either the neutralino or gravitino being the LSP.  We make use of the ATLAS direct SUSY searches at the 7/8 TeV LHC as well as
expected results from 14 TeV operations, and the
anticipated precision measurements of the Higgs Boson couplings at the 14 TeV LHC and at the ILC.  
We demonstrate that the future Higgs coupling determinations can deeply probe the pMSSM parameter space and, in particular, can observe the effects of models that are projected to
evade the direct searches at the 14 TeV LHC with 3 ab$^{-1}$ of integrated luminosity.  In addition,
we compare the reach of the Higgs coupling determinations to the direct heavy Higgs searches in the $M_A - \tan\beta$ plane
and show that they cover orthogonal regions.  This analysis
demonstrates the complementarity of the direct and indirect approaches in searching for Supersymmetry, and the importance of precision studies of the properties of the
Higgs Boson.

\end{abstract}

\section{Overview}

Almost 50 years since its theoretical inception, 
the Higgs boson has been discovered at the LHC{\cite{higgs}}.  Nonetheless, the Higgs boson remains a mystery, and its discovery has unlocked many questions
about its nature that are related to its special role in the Universe.   
Now that the Higgs boson mass is known, the Standard Model (SM) predicts its interactions and properties with no free parameters.  Any deviation 
from these predictions provides unambiguous evidence for new physics, making a rigorous study of the Higgs a focus of upcoming operations at the LHC, as well as at
future colliders.  This quest to determine the properties of the Higgs goes hand in hand with direct searches for new physics at the LHC.  In particular, it is
crucial to understand how the two modes of exploration are intertwined.  In this paper, we examine this connection within the framework of the 
p(henomenological) MSSM{\cite{Djouadi:1998di}}.

The pMSSM provides an excellent structure for a systematic and comprehensive survey of constraints on Supersymmetry (SUSY) and for the investigation of complementary
approaches to detecting its existence.  Towards this end, we have previously embarked on a detailed study of signatures 
for the pMSSM at the 7, 8 and 14 TeV LHC~{\cite {us1,us2,usnew}} and have compared the LHC search reach to that of searches for dark matter via
direct and indirect detection~\cite{Cahill-Rowley:2013dpa}.  Our focus on Supersymmetry stems from its attractiveness as a candidate for new physics.
Its presence at the weak-scale would stabilize the Higgs sector under quantum corrections, provide a natural thermal dark matter candidate, and accommodate
unification of the gauge couplings.

The pMSSM is the most general version of the R-parity conserving MSSM when it is subjected to a minimal set 
of experimentally-motivated guiding principles: ($i$) CP conservation, ($ii$) Minimal Flavor Violation at the electroweak scale so that flavor 
physics is controlled by the CKM mixing matrix and the Yukawa couplings of the SM fermions, ($iii$) degenerate 1\textsuperscript{st} and 
2\textsuperscript{nd} generation sfermion masses. In addition, it is assumed that ($iv$) the Yukawa couplings and A-terms for the first two 
generations can be safely neglected.  In particular, theoretical assumptions about physics at high scales, {\it e.g.}, the nature of SUSY 
breaking, are absent in order to capture electroweak scale phenomenology for which a UV-complete theory may not yet exist. Imposing these principles 
decreases the number of free parameters in the MSSM at the TeV-scale from 105 to 19 for the case of a neutralino LSP.  If the gravitino 
mass is included so that it plays the role of the LSP, an additional parameter is required.  

With respect to the production of new physics at an accelerator, a key question is whether its signature can be detected  
given our understanding of the backgrounds arising from SM processes (provided the new particles are kinematically accessible). In 
particular, it is important to determine how experimental analyses can probe the full parameter space of 
interest within any specific model. This is certainly true in the case of Supersymmetry. However, even in the simplest SUSY scenario, the MSSM, the number of free 
parameters ($\sim$ 100) is much too large to study in complete generality. A traditional approach is to assume the existence of a UV-complete 
theory with minimal set of parameters (such as mSUGRA{\cite {SUSYrefs}) from which the properties of the sparticles at the TeV scale can be determined and studied in 
detail. While such an approach is often quite valuable~\cite{Cohen:2013kna}, these scenarios can be phenomenologically 
limiting and many are under increasing tension with a wide range of experimental data, including the $\sim 126$ GeV mass of the Higgs. At the 
opposite end of the spectrum, simplified model scenarios can be employed to estimate constraints from the LHC, thereby bounding the model parameter space in a 
process-by-process fashion. However, a concern in this case is that the simplified models are not capturing the `big picture' of what is occurring in the full 
underlying theory.  The more general pMSSM circumvents the limitations of these other approaches.  
The increased dimensionality of the parameter space not only allows for a somewhat less prejudiced study of SUSY, 
but also yields valuable information on unusual scenarios, identifies weaknesses in the LHC analyses, and can be used to combine 
results from many individual and independent SUSY related searches. 

To study the pMSSM, we generate large sets of models by randomly scanning the parameter space.
The 19/20 parameters and the ranges of values that we employ in our scans are listed in Table~\ref{ScanRanges}. In order to sample the pMSSM space as 
thoroughly as possible, we generate many millions of model points (using SOFTSUSY{\cite{Allanach:2001kg}} and checking for consistency with 
SuSpect{\cite{Djouadi:2002ze}}), with each point corresponding to a specific set of values for the parameters.  We then subject these individual `models'  
to a global set of collider, flavor, precision, dark matter and theoretical constraints~\cite{us1}.  In particular,
we do not assume that the LSP relic density necessarily saturates the WMAP/Planck value{\cite{omega}}, $\Omega h^2 \simeq 
0.12$,  in order to allow for the possibility of multi-component DM. (For example, the axions introduced to solve the strong CP problem may contribute significantly to the DM relic density.) 
Roughly  $\sim$225k models for each type of LSP survive this initial selection and can be used for further physics studies. Decay patterns of the 
SUSY partners and the extended Higgs sector are calculated using a privately modified version of SUSY-HIT{\cite{Djouadi:2006bz}} as well as the 
most recent version of HDECAY{\cite {HDECAY}}. Since our scan ranges include sparticle masses up to 4 TeV, an upper limit chosen to enable phenomenological studies at the 14 TeV LHC, the majority of neutralinos and charginos are nearly pure electroweak eigenstates.  This is due to the off-diagonal elements of the corresponding mass matrices 
being at most $M_W$. This has important 
implications for the resulting collider and DM phenomenology\cite{us1,us2,usnew,Cahill-Rowley:2013dpa}. We note that both of these model sets were generated before the Higgs boson was discovered. For the neutralino (gravitino) model set 
we find that roughly $\simeq 20 (10)\%$ of the models are found to satisfy $m_h=126\,\pm\,3$ GeV; clearly, we will focus on these subsets in the analyses that 
follow.

In addition to these two large pMSSM model sets, we have also generated a third, somewhat smaller, specialized set of 
`natural' models with the neutralino being identified as the LSP.  These models predict $m_h=126\,\pm\,3$ GeV, have an LSP that {\it does} saturate the 
WMAP/Planck relic density, and yield values 
of fine-tuning (FT) better than $1\%$ employing the traditional Ellis-Barbieri-Giudice measure~\cite{ebg}. This low-FT model 
set will also be included as part of the present study. 
In order to produce this model 
set, we modified the parameter scan ranges as indicated in Table~\ref{ScanRanges} to greatly increase the likelihood of achieving both low FT and 
a thermal relic density in the desired range. In addition to these modified scan ranges, we also required $|M_1/\mu|<1.2$ and $|X_t|/m_{\tilde t} >1$, where
$X_t=A_t-\mu\cot\beta$ quantifies the mixing between the stop-squarks with $m_{\tilde t}$ being the geometric mean of the tree-level stop masses.
Amongst other things, this requires a 
bino-like LSP, light Higgsinos and highly mixed stops. We generated $\sim 3.3 \times 10^8$ low-FT points in this 19-dimensional parameter 
space and required consistency with current precision, flavor, DM and collider constraints as before. 
Due to the difficulty of satisfying this set of requirements, only $\sim$ 10.2k low-FT models were found to be viable for further study. 

Within each pMSSM model, the characteristics of the lightest CP-even Higgs, $h$, as well as the entire superpartner spectrum, are calculable (to several loops) from
the chosen values of the soft-breaking parameters in the underlying Lagrangian.
Given this correspondence, we can address the connection between the predicted SUSY Higgs properties and the direct searches for SUSY at the LHC.  In particular,
we seek to address two questions: ($i$) 
How will potentially null searches for SUSY at the LHC influence the predicted properties of the Higgs boson? ($ii$) What can be learned about the properties of the superpartners
from precision measurements of the Higgs Boson couplings? 

In what follows, we briefly discuss the impact on our model sets of the 7, 8, direct SUSY searches at the LHC, 
as well as the expectations for 14 TeV searches in Section 2.  In Section 3 we examine the predictions of the properties of the lightest Higgs in the pMSSM.
We discuss the impact of measurements of the Higgs properties on the pMSSM from current data and from future measurements at the 14 TeV LHC with 300 fb$^{-1}$ and 3 ab$^{-1}$,
as well as the proposed International Linear Collider (ILC), in Section 4.  Our conclusions are given in Section 5.

\begin{table}
\centering
\begin{tabular}{|c|c|c|} \hline\hline
Parameter & General Neutralio/Gravitino Set & Low Fine-Tuned Set \\
\hline\hline
$m_{\tilde L(e)_{1/2,3}}$ & $100 \gev - 4 \tev$ & 100\gev - 4\tev \\ 
$m_{\tilde Q(u,d)_{1/2}}$ & $400 \gev - 4 \tev$ & 100\gev - 4\tev  \\ 
$m_{\tilde Q(u,d)_{3}}$ &  $200 \gev - 4 \tev$ &  100\gev - 4\tev\\
$|M_1|$ & $50 \gev - 4 \tev$ & 25\gev - 552\gev \\
$|M_2|$ & $100 \gev - 4 \tev$ &  100\gev - 2.1\tev \\
$|\mu|$ & $100 \gev - 4 \tev$ & 100\gev - 460\gev \\ 
$M_3$ & $400 \gev - 4 \tev$ &  400\gev - 4\tev \\ 
$|A_{t,b,\tau}|$ & $0 \gev - 4 \tev$ & 0\gev - 2.3\tev ($A_t$ only) \\ 
$M_A$ & $100 \gev - 4 \tev$ &  100\gev - 4\tev\\ 
$\tan \beta$ & $1 - 60$ & 1 - 60 \\
$m_{3/2}$ & 1 eV$ - 1 \tev$ ($\tilde{G}$ LSP) & - \\
\hline\hline
\end{tabular}
\caption{Scan ranges for the 19 (20) parameters of the pMSSM with a neutralino (gravitino) LSP. The gravitino mass is scanned with 
a log prior. All other parameters are scanned with flat priors; we expect this choice to have little qualitative impact on 
our results for observables~\cite{Djouadi:1998di}.}
\label{ScanRanges}
\end{table}

\section{LHC SUSY Searches}

To begin this study, we first ascertain which models in each of our three sets are excluded at the 7,8 TeV LHC, and which can be probed at 14 TeV. 
Once these current constraints and future expectations for the pMSSM parameter space are characterized, we can determine how the properties of
the lightest SUSY Higgs boson are affected by the direct searches, and quantify how they may differ from SM predictions.  Such correlations between 
the direct search results and the properties of the Higgs can address the questions posed above. 

We begin this step of the analysis with a brief overview of our procedure for computing the effects of the LHC direct SUSY searches on the pMSSM.
In general, we replicate the suite of ATLAS SUSY search analyses as closely as possible employing fast Monte Carlo.  We also include 
several searches performed by CMS. The specific analyses applied to our pMSSM model sets are briefly summarized in 
Tables~\ref{SearchList7} and~\ref{SearchList8}. We augment the standard MET-based SUSY channels by including searches for heavy stable charged particles and 
a heavy neutral SUSY Higgs decaying into $\tau^+\tau^-$ as performed by CMS~\cite{CMSextra}, as well as measurements of the rare decay mode $B_s\to \mu^+\mu^-$ 
as discovered by CMS and LHCb~\cite{BSMUMU}.  All of these play distinct and important roles in covering the pMSSM parameter space.  Details of our
analysis and results are discussed at length in our previous work~\cite{us1,us2,usnew}, with the most recent description of
our final results for 7 and 8 TeV and expectations for 14 TeV given in \cite{usnew}.  Here, we provide a concise summary of the salient features of this work
in order to provide a basis for investigating the properties of the Higgs.

\begin{table}
\centering
\begin{tabular}{|l|l|c|c|c|} \hline\hline
Search & Reference & Neutralino & Gravitino & Low-FT   \\
\hline
2-6 jets & ATLAS-CONF-2012-033  & 21.2\% &  17.4\% & 36.5\% \\
multijets & ATLAS-CONF-2012-037 & 1.6\%  & 2.1\% & 10.6\% \\
1 lepton & ATLAS-CONF-2012-041 & 3.2\%  & 5.3\% & 18.7\%  \\

HSCP      &  1205.0272  & 4.0\% & 17.4\% & $<$0.1\%  \\
Disappearing Track  & ATLAS-CONF-2012-111 & 2.6\%  & 1.2\% & $<$0.1\% \\
Muon + Displaced Vertex  & 1210.7451 & - & 0.5\% & - \\
Displaced Dilepton & 1211.2472 & - & 0.8\% & - \\

Gluino $\to$ Stop/Sbottom   & 1207.4686 & 4.9\% &  3.5\% & 21.2\% \\
Very Light Stop  & ATLAS-CONF-2012-059 & $<$0.1\% & $<$0.1\% & 0.1\%  \\
Medium Stop  & ATLAS-CONF-2012-071 & 0.3\% & 5.1\% & 2.1\% \\
Heavy Stop (0$\ell$)  & 1208.1447 & 3.7\% & 3.0\% & 17.0\% \\
Heavy Stop (1$\ell$)   & 1208.2590 & 2.0\% & 2.2\% & 12.6\% \\
GMSB Direct Stop  & 1204.6736 & $<$0.1\% & $<$0.1\% & 0.7\% \\
Direct Sbottom & ATLAS-CONF-2012-106 & 2.5\% & 2.3\% & 5.1\% \\
3 leptons & ATLAS-CONF-2012-108 & 1.1\% & 6.1\% & 17.6\% \\
1-2 leptons & 1208.4688 & 4.1\% & 8.2\% & 21.0\% \\
Direct slepton/gaugino (2$\ell$)  & 1208.2884 & 0.1\% & 1.2\% & 0.8\% \\
Direct gaugino (3$\ell$) & 1208.3144 & 0.4\% & 5.4\% & 7.5\% \\
4 leptons & 1210.4457 & 0.7\% & 6.3\% & 14.8\% \\
1 lepton + many jets & ATLAS-CONF-2012-140 & 1.3\% & 2.0\% & 11.7\% \\
1 lepton + $\gamma$ & ATLAS-CONF-2012-144 & $<$0.1\% & 1.6\% & $<$0.1\% \\
$\gamma$ + b & 1211.1167 & $<$0.1\% & 2.3\% & $<$0.1\% \\
$\gamma \gamma $ + MET & 1209.0753 & $<$0.1\% & 5.4\% & $<$0.1\% \\

$B_s \to \mu \mu$ & 1211.2674 & 0.8\% & 3.1\% & * \\
$A/H \to \tau \tau$ & CMS-PAS-HIG-12-050 & 1.6\% & $<$0.1\% & * \\

\hline\hline
\end{tabular}
\caption{7 TeV LHC searches included in the present analysis, and the corresponding fraction of the neutralino, gravitino and low-FT pMSSM 
model sets excluded by each channel. Note that in the case of the last two rows the experimental constraints have already been included 
in the model generation process for the low-FT model set.}
\label{SearchList7}
\end{table}

\begin{table}
\centering
\begin{tabular}{|l|l|c|c|c|} \hline\hline
Search & Reference & Neutralino & Gravitino & Low-FT    \\
\hline

2-6 jets   & ATLAS-CONF-2012-109 & 26.7\% & 22.5\% & 44.9\% \\
multijets   & ATLAS-CONF-2012-103 & 3.3\% & 5.6\% & 20.9\% \\
1 lepton     & ATLAS-CONF-2012-104 & 3.3\% & 6.0\% & 20.9\% \\
SS dileptons & ATLAS-CONF-2012-105 & 4.9\% & 12.5\% & 35.5\% \\
2-6 jets   & ATLAS-CONF-2013-047 & 38.0\% & 31.1\% & 56.5\% \\

HSCP      &  1305.0491  & - & 23.0\% & -  \\

Medium Stop (2$\ell$) & ATLAS-CONF-2012-167 & 0.6\% & 8.1\% & 4.9\% \\
Medium/Heavy Stop (1$\ell$) & ATLAS-CONF-2012-166 & 3.8\% & 4.5\% & 21.0\% \\
Direct Sbottom (2b) & ATLAS-CONF-2012-165 & 6.2\% & 5.1\% & 12.1\% \\
3rd Generation Squarks (3b) & ATLAS-CONF-2012-145 & 10.8\% & 9.9\% & 40.8\% \\
3rd Generation Squarks (3$\ell$) & ATLAS-CONF-2012-151 & 1.9\% & 9.2\% & 26.5\% \\
3 leptons & ATLAS-CONF-2012-154 & 1.4\% & 8.8\% &32.3\% \\
4 leptons & ATLAS-CONF-2012-153 & 3.0\% & 13.2\% & 46.9\% \\
Z + jets + MET & ATLAS-CONF-2012-152 & 0.3\% & 1.4\% &6.8\% \\

\hline\hline
\end{tabular}
\caption{Same as in the previous table but now for the 8 TeV ATLAS MET-based SUSY searches. Note that when all the channels from this table and the previous table are 
combined,  we find that $\sim 45.5~(61.3,~74.0)\%$ of these models are excluded by the LHC for the neutralino 
(gravitino, low-FT) model set.}
\label{SearchList8}
\end{table}

Briefly stated, our procedure is as follows: We generate SUSY events for each model for all relevant (up to 85) production channels with PYTHIA 
6.4.26~\cite{Sjostrand:2006za}, and then pass the events through fast detector simulation using PGS 4~\cite{PGS}. Both programs have been modified to, 
{\it e.g.}, correctly deal with gravitinos, multi-body decays, hadronization of stable colored sparticles, and ATLAS b-tagging. We then scale our event rates to NLO by computing the relevant K-factors using Prospino 2.1~\cite{Beenakker:1996ch}. The 
individual searches are then implemented using our customized analysis code{\cite {us}}, which follows the published experimental cuts and selection criteria as 
closely as possible. This analysis code is validated for each of the many search regions for every channel, employing the benchmark model points 
provided by ATLAS (and CMS). Models are then excluded using the 95\% $CL_s$ limits as employed by ATLAS (and CMS). For the purpose of obtaining
the direct SUSY search results on the two large model sets, we
perform this analysis {\it without} requiring the Higgs mass constraint, $m_h=126\,\pm\,3$ GeV (combined experimental and theoretical errors) so that we 
can understand its influence on the search results. Recall that roughly $\sim 20 (10)\%$ of models in the neutralino(gravitino) model set predict a Higgs mass in 
this range. While we observe some variation amongst the individual searches, we find that once the channels are combined, the overall pMSSM model
coverage is to an excellent approximation {\it independent} of the value of the Higgs mass  
\cite{usnew}. Conversely, the fraction of neutralino and gravitino LSP models predicting the observed Higgs mass is also found to be approximately 
independent of whether or not the direct SUSY search results have been enforced. This result is very powerful and demonstrates the approximate decoupling of the direct 
SUSY search results from the mass of the Higgs boson. Of course, for this study, in which we specifically examine the properties of the 
Higgs boson itself, we restrict our investigation to the subset of the neutralino and gravitino LSP model samples that predict $m_h=126\,\pm\,3$ GeV. No additional 
requirements on the Higgs mass are necessary for the low-FT set, since in this case the Higgs mass constraint is imposed during the model generation 
process.   

Tables~\ref{SearchList7} and~\ref{SearchList8} also show 
the coverage of our pMSSM model sets from the 7 and 8 TeV search constraints.  We find that $\sim 45.5 (61.3, 74.0)\%$
of the neutralino (gravitino, low-FT) model samples are excluded by the LHC.  In particular, we find that numerous models with light squarks
and gluinos (500-1000 GeV) are currently viable.  These results demonstrate that much phase-space is left to
accommodate natural Supersymmetry.

In addition to the searches performed at 7 and 8 TeV, future LHC operations at $\sim 14$ TeV will greatly extend the coverage of the pMSSM parameter space.
For our 14 TeV analysis, we considered the impact of two of the most powerful searches to be performed by ATLAS~\cite{atlas14}, namely the zero-lepton jets +MET
and the zero- and one-lepton stop channels.  We have simulated these channels~\cite{usnew} in a manner identical to that described above
for the 7 and 8 TeV searches.  We have extrapolated the results expected by ATLAS at 300 fb$^{-1}$ of integrated luminosity to 3 ab$^{-1}$ by scaling
the required signal rate.  Due to the large CPU required to generate events at these luminosities, we restricted our study to the subset of models that
remain viable after the 7,8 TeV constraints and predict the observed Higgs mass.  
We find that with 300 (3000) fb$^{-1}$ of data, the combination of these searches covers 90.83\% (97.15\%) of the
neutralino LSP model set, 83.22\% (93.29\%) for the gravitino LSP model set,
and 97.69\% (100\%) of the low-FT model sample.
Clearly, the 14 TeV LHC will provide a more definitive statement on the existence of natural Supersymmetry, even in complex forms such as the pMSSM,
and the discovery space of the upcoming run is significant.

These results of the direct LHC SUSY searches will be employed below in our study of the Higgs couplings.

\section{Determination of Higgs Properties}

In this section, we show how the pMSSM parameter space can be constrained by
the measured properties of the Higgs. For this analysis, we must first
determine the extent that the couplings of the light CP-even Higgs boson in the pMSSM differ from the expectations for the SM Higgs, and then we can compare these results 
to the current and expected future experimental determinations of the couplings. We make several such comparisons corresponding to the anticipated evolution 
of our knowledge about the allowed values of the Higgs couplings: ($i$) current data{\cite {now}}, ($ii$) measurements that are expected to be attainable 
at the 14 TeV LHC with an integrated luminosity of 0.3(3) ab$^{-1}${\cite {14TeVLHC}}, and finally ($iii$) projected measurements at the ILC with two different run plans being
250 \infb\ at 250 GeV plus 250 \infb\ at 500 GeV, as well as an upgrade to 1150 \infb\ at 250 GeV plus 1600 \infb\ at 500 GeV plus 2500 \infb\ at 1000 GeV center of mass 
energy~{\cite {ILC}}. 

To calculate the Higgs couplings in the pMSSM and in the SM, we employ HDECAY 5.11.
We note that since the full set of computed SUSY loop corrections for the $h \to WW$ and $h\to ZZ$ partial 
widths are not yet incorporated in HDECAY, we unfortunately can not employ these very important modes to constrain our pMSSM model sample. We follow the standard approach, using the narrow width approximation (NWA) and defining the signal strength for a given production channel (e.g. $gg,VBF\to h$), with the subsequent decay into the
final state, $h \to X$, normalized to the corresponding SM value, as 
\begin{equation}
\mu_{gg,VBF}(X) = {{\sigma(gg,VV\to h)~B(h \to X)}\over {SM}}\,.
\end{equation}
For final states that do not involve the top quark, we can also define the ratio of the squares of the couplings to 
their corresponding SM values by simply forming the ratio of the relevant partial decay widths,
\begin{equation}
r_X = {{\Gamma (h \to X)}\over {SM}}\,,
\end{equation}
for the final states $X=ZZ,~W^+W^-,~\bar b b, ~\bar c c, ~\tau^+\tau^-,~gg, ~\gamma\gamma, ~\gamma Z$.  The case of the $ht\bar t$ coupling must be handled 
separately and can only be directly accessed via associated $t\bar th$ production. We are, of course, also interested in the branching fraction 
for Higgs decays into the lightest neutralino{\footnote {The LEP limits of $\sim$100 GeV on the mass of charged sparticles, which we apply strictly, constrain the possible invisible decay modes of the Higgs. We note that neutral winos, Higgsinos and sneutrinos are required to have a charged partner with a similar mass, thus preventing them from being decay products of the Higgs.}},
producing a final state which is purely invisible or accessed by jets+MET, depending on the production channel.
Searches for invisible decays into the LSP are very interesting because of their potential to place significant constraints on 
the SUSY parameter space, particularly when results from ILC500 are employed, as we shall see below.

\begin{figure}[htbp]
\centerline{\includegraphics[width=3.5in]{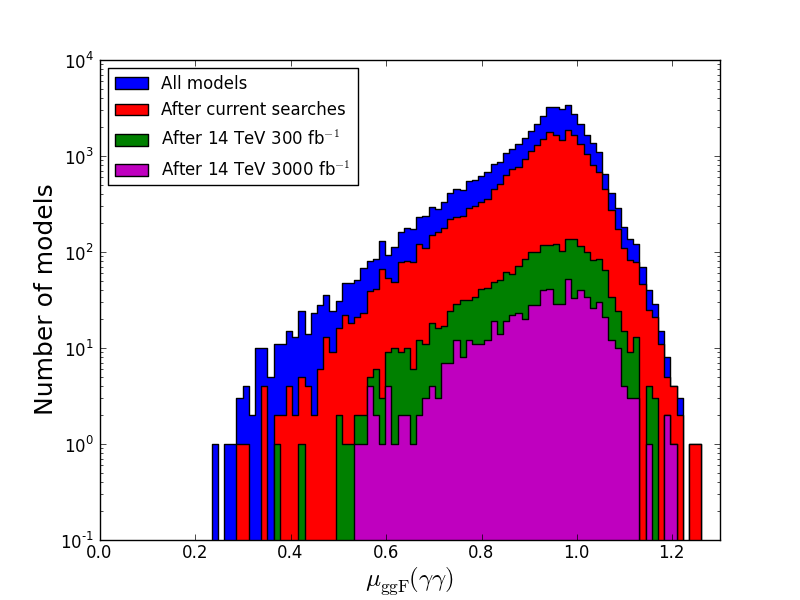}
\hspace{-0.50cm}
\includegraphics[width=3.5in]{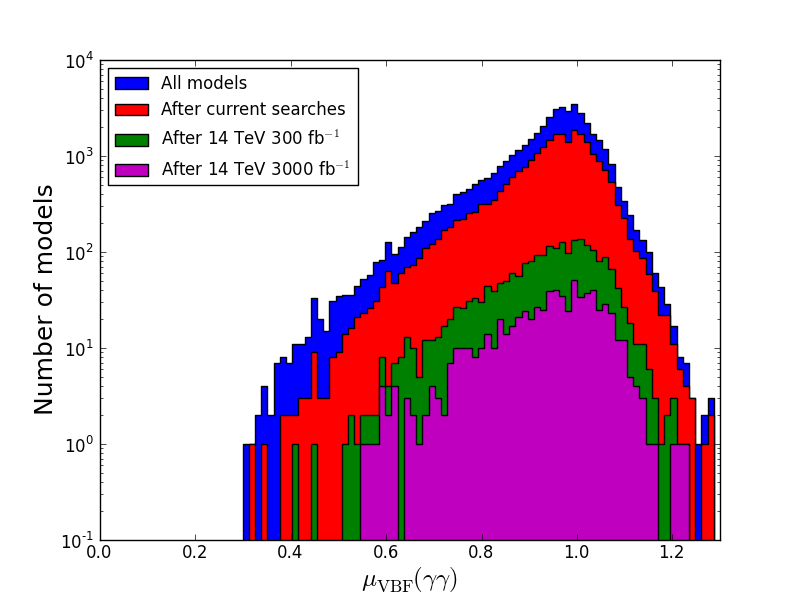}}
\vspace*{0.50cm}
\centerline{\includegraphics[width=3.5in]{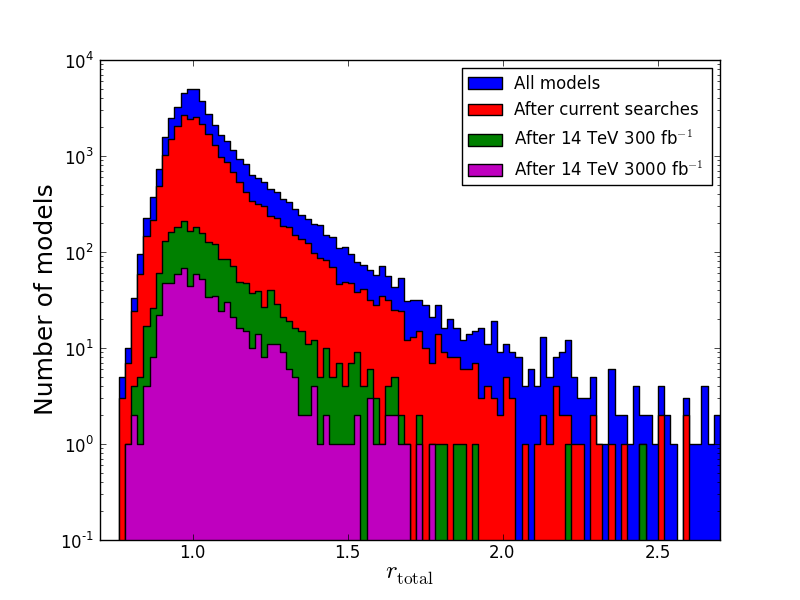}}
\vspace*{-0.10cm}
\caption{Histograms of signal strengths for $h \to \gamma \gamma$ in the $gg$-fusion (top left) and vector boson fusion (top right) production channels for the subset 
of neutralino models that predict $m_h=\,126\pm\,3$ GeV. The blue (red) histogram represents models before any ATLAS searches (after the 7 and 8 
TeV SUSY searches) are applied, while the green (purple) histograms show models that are expected to survive the zero-lepton jets plus MET plus the 0,1-$\ell$ stop searches 
at 14 TeV, assuming a luminosity of 300 (3000) fb$^{-1}$. The ratio $r_{total}$ for the total width of the Higgs is analogously shown in the bottom panel.}
\label{figA}
\end{figure}

To get an initial understanding of the distribution of Higgs properties in the various pMSSM model sets, it is instructive to first study a few examples. Figure~\ref{figA}, 
shows the distribution of the $h\to \gamma\gamma$ signal strength for both the $gg$-fusion and vector boson fusion production channels in the neutralino 
LSP model set (with $m_h=126\,\pm\,3$ GeV), along with the effect of the current 7/8 TeV and future 14 TeV ATLAS searches on this 
distribution~\cite{usnew} as indicated. Other than the obvious fact that these distributions peak near unity but have long tails, 
the most important observation is that the shape of these distributions (up to statistical fluctuations) is essentially unaffected by the 
imposition of the ATLAS direct SUSY searches. Furthermore, the shape of the distribution for the ration of total widths, $r_{total}= \Gamma(h\to All)/SM$, for the neutralino models  
demonstrates that this shape invariance is maintained for the other observables.  We therefore see that SUSY searches and Higgs boson properties are to a very good 
approximation `orthogonal'. As we will show below, the other final states exhibit a similar behavior, answering our first question above: 
Future null direct SUSY searches at the LHC will, to a good approximation (as is seen here except for statistical limitations), 
not significantly modify the range of values that we expect for the SUSY Higgs 
couplings.

\begin{figure}[htbp]
\centerline{\includegraphics[width=5.5in]{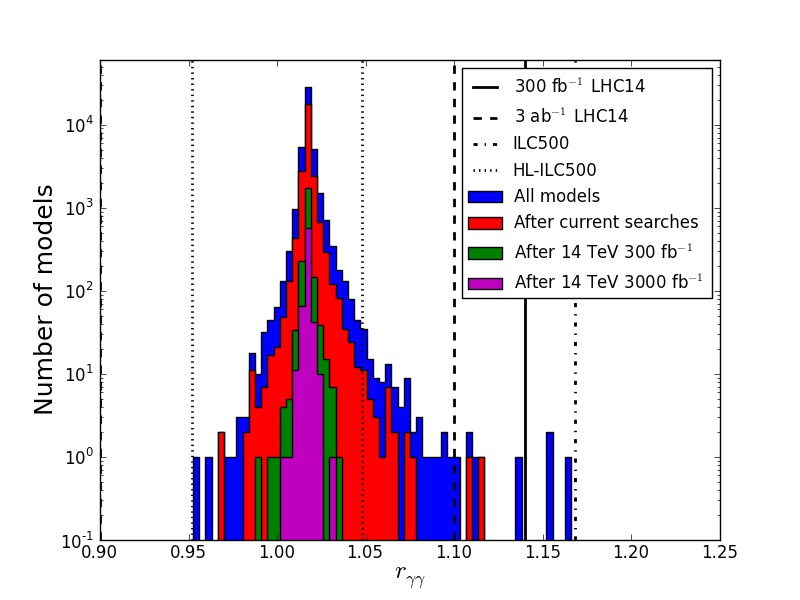}}
\vspace*{0.50cm}
\centerline{\includegraphics[width=3.5in]{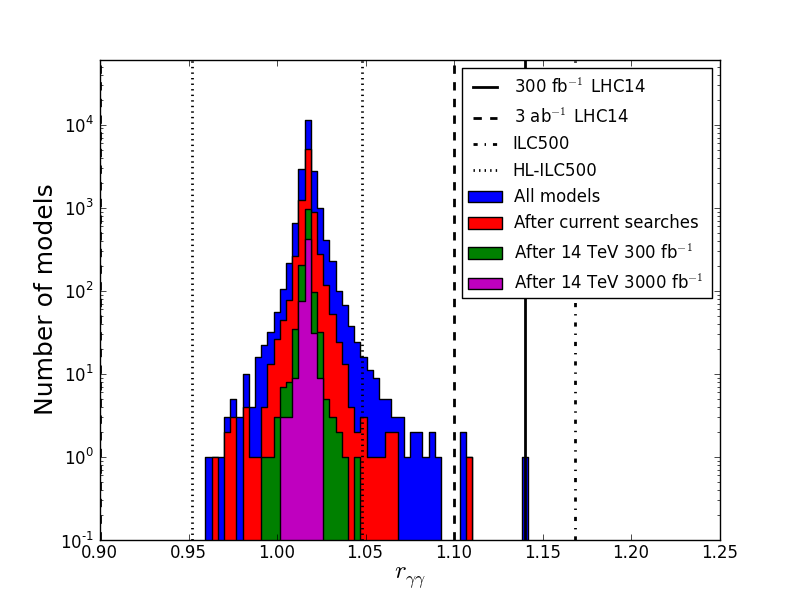}
\hspace{-0.50cm}
\includegraphics[width=3.5in]{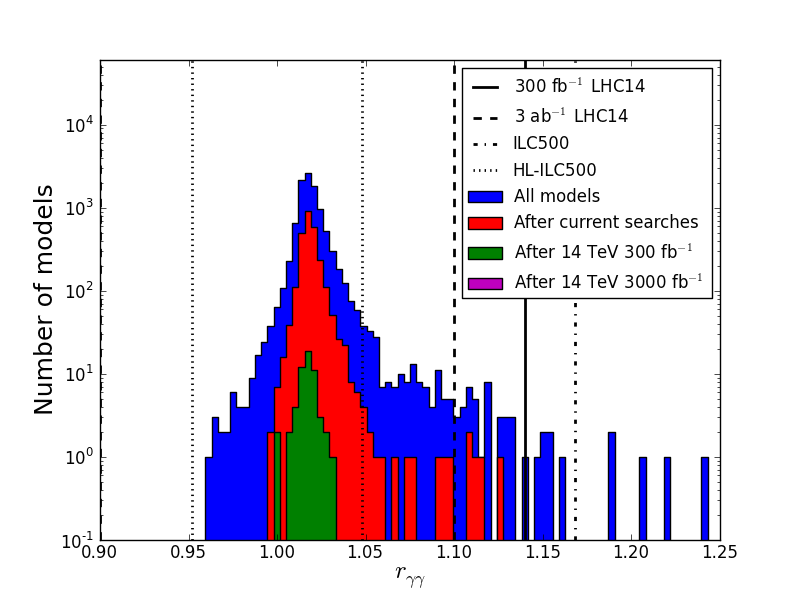}}
\vspace*{-0.10cm}
\caption{Histograms of the ratio of partial widths for $h \to \gamma \gamma$ for the subset of neutralino (top), gravitino (lower left) and low-FT models (lower 
right) that predict $m_h=126\,\pm\,3$ GeV. The blue (red) histogram represents models before any ATLAS searches (after the 7 and 8 TeV SUSY searches) are applied 
while the green (purple) histograms show models that are expected to survive both the zero-lepton jets plus MET and the 0,1-$\ell$ stop searches at 14 TeV, assuming 
an integrated luminosity of 300 (3000) fb$^{-1}$. The vertical lines show the expected future limits on $r_{\gamma \gamma}$, and are discussed in the text.}
\label{figB}
\end{figure}
\begin{figure}[htbp]
\centerline{\includegraphics[width=5.5in]{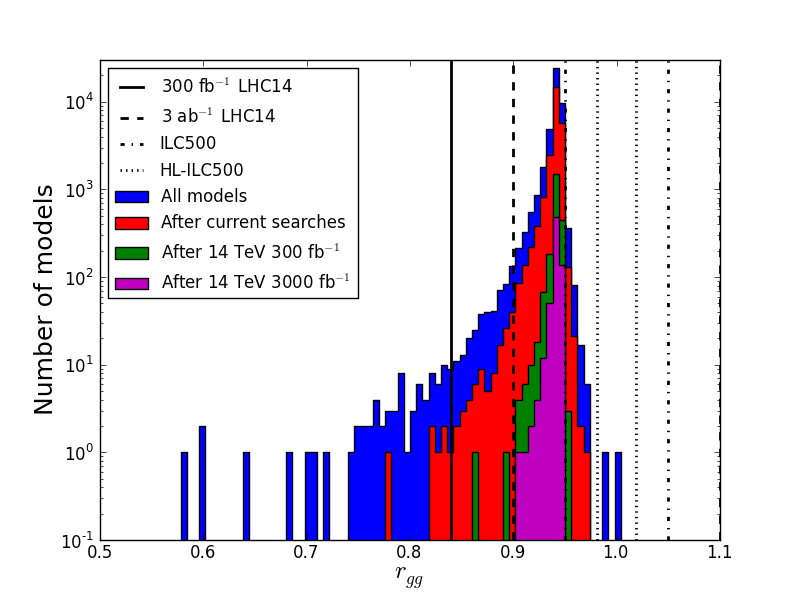}}
\vspace*{0.50cm}
\centerline{\includegraphics[width=3.5in]{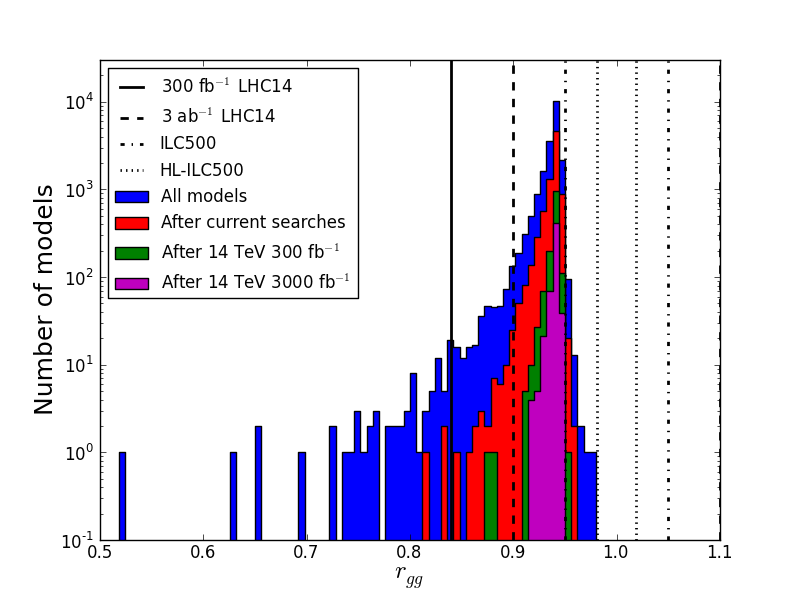}
\hspace{-0.50cm}
\includegraphics[width=3.5in]{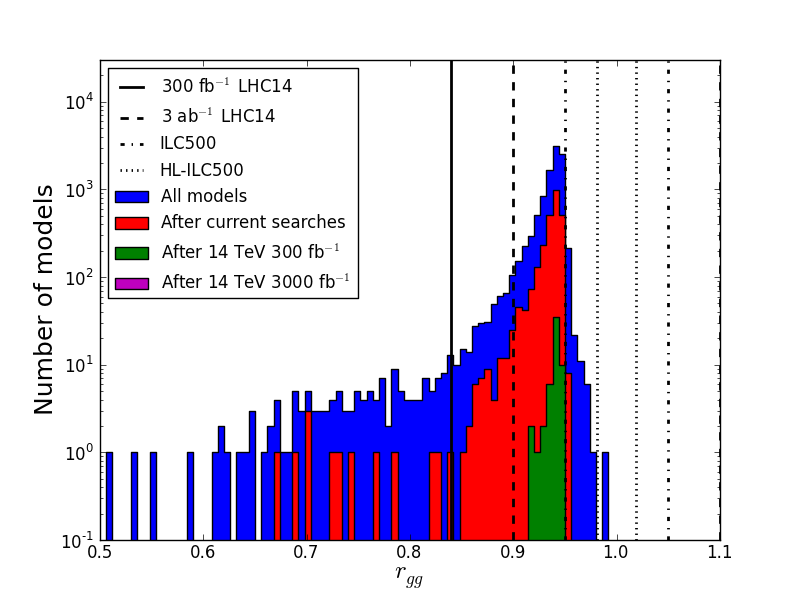}}
\vspace*{-0.10cm}
\caption{Same as the previous Figure but now for $h\to gg$.}
\label{figC}
\end{figure}
\begin{figure}[htbp]
\centerline{\includegraphics[width=5.5in]{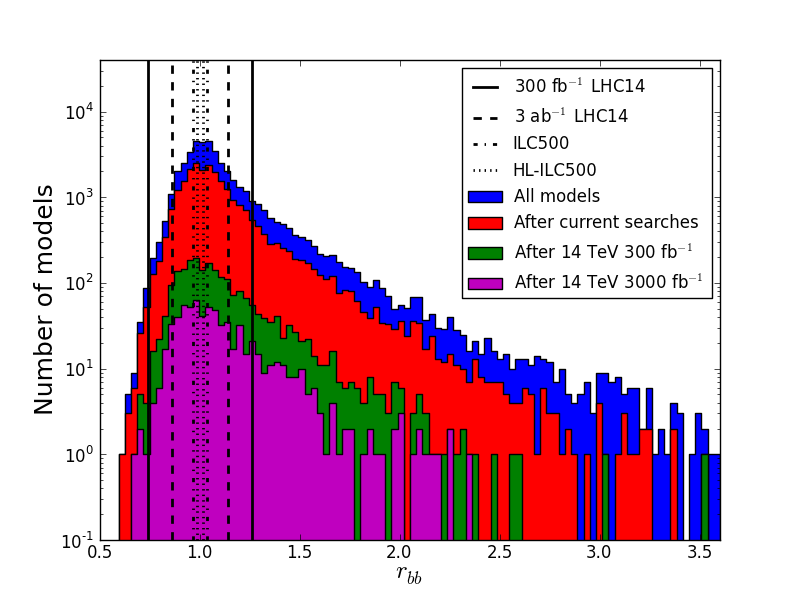}}
\vspace*{0.50cm}
\centerline{\includegraphics[width=3.5in]{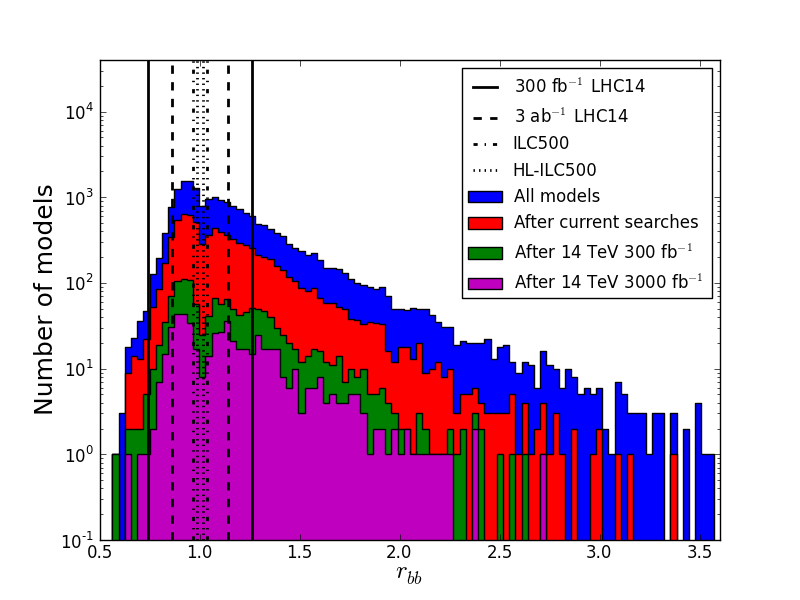}
\hspace{-0.50cm}
\includegraphics[width=3.5in]{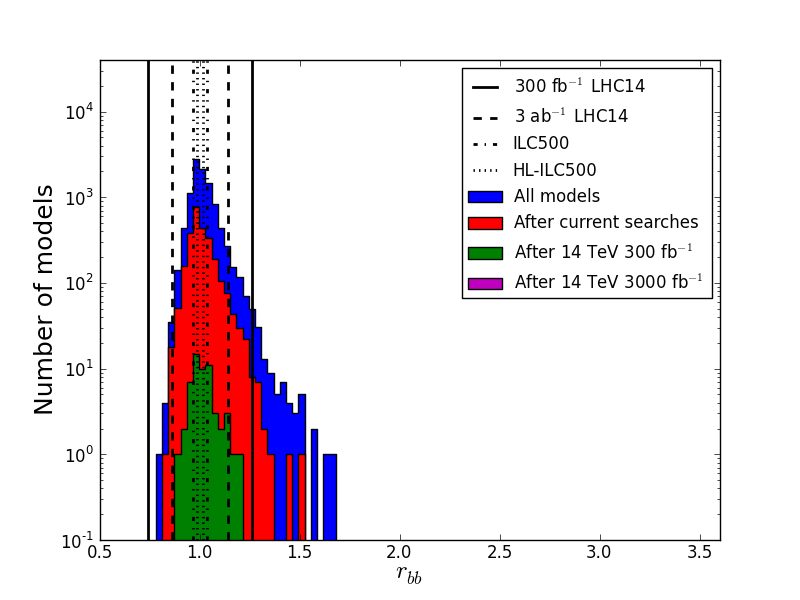}}
\vspace*{-0.10cm}
\caption{Same as in Figure~\ref{figB} but now for $h\to b \bar b$.}
\label{figD}
\end{figure}
\begin{figure}[htbp]
\centerline{\includegraphics[width=5.5in]{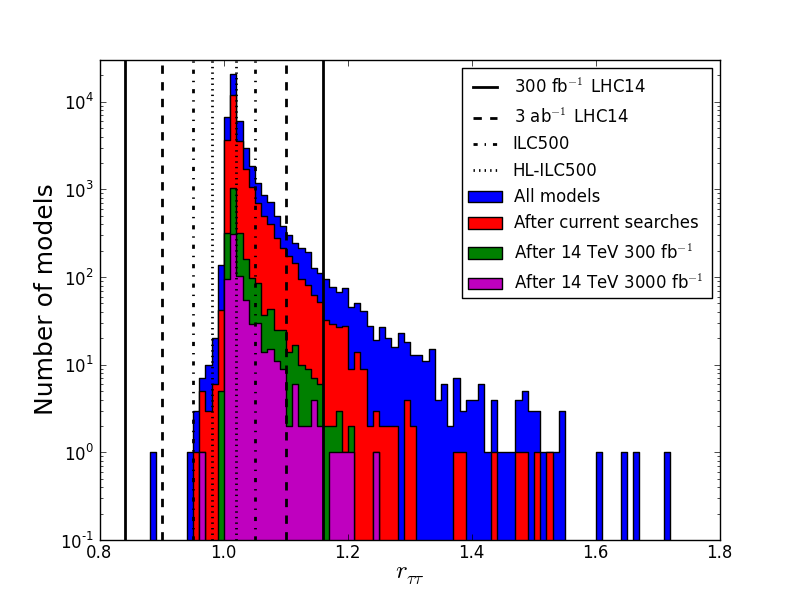}}
\vspace*{0.50cm}
\centerline{\includegraphics[width=3.5in]{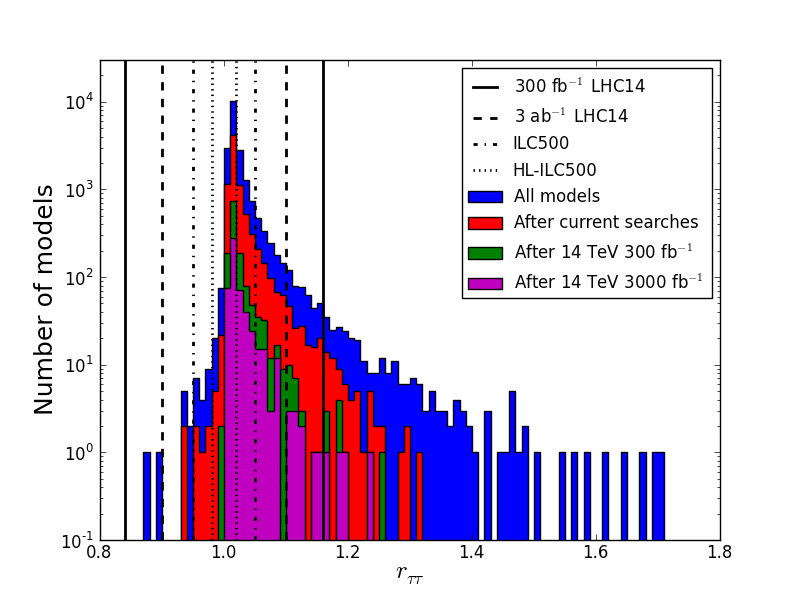}
\hspace{-0.50cm}
\includegraphics[width=3.5in]{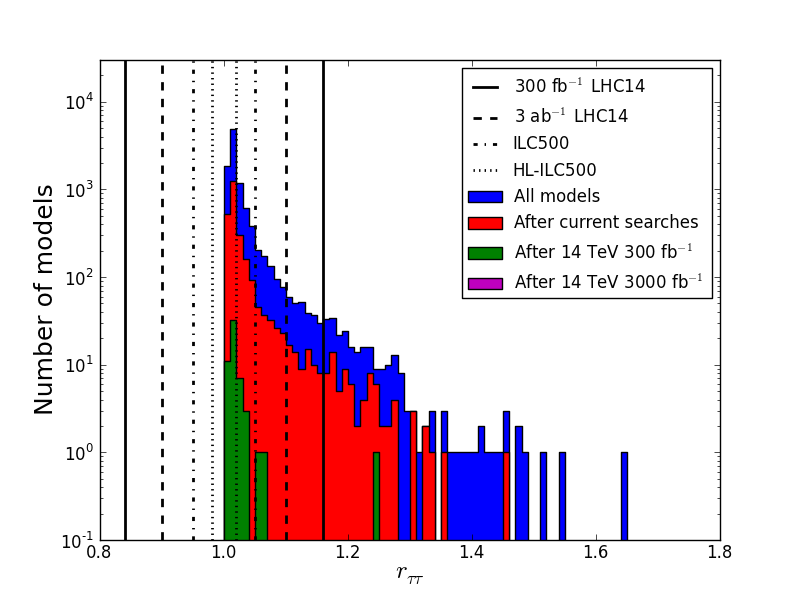}}
\vspace*{-0.10cm}
\caption{Same as in Figure~\ref{figB} but now for $h\to \tau^+\tau^-$.}
\label{figE}
\end{figure}

We now turn our attention to the predicted distributions for the values of the various partial width distributions, $r_X$, in each model set, and the 
effect of the future LHC direct SUSY searches on these distributions. We will return to these distributions later in our subsequent analysis to understand the effects 
of the future Higgs coupling measurements.

Figure~\ref{figB} shows a histogram for the ratio $r_{\gamma\gamma}$ in the three different model sets. The 
vertical lines appearing in these plots are discussed in detail in the next section, and represent the anticipated Higgs coupling measurement sensitivities provided by future measurements 
at the 14 TeV LHC and ILC500 as discussed above and indicated in the figure. Qualitatively, we see that the effect of the LHC direct
searches on the $r_{\gamma\gamma}$ distributions is to decrease the normalization 
while preserving the overall shapes of the distributions for all three model sets. Deviations from this general behavior are mainly seen in the tails of the 
distributions, where the statistics are low. The different responses of each model set to the direct LHC SUSY searches can be seen by observing the  
differing impacts of the searches on the distribution areas. Interestingly, the $r_{\gamma\gamma}$ distribution in the neutralino model 
set has a very different shape compared with the corresponding distribution of diphoton signal strengths shown in Figure~\ref{figA}, which are coupled to the production
channels; this difference results mainly from 
large corrections to the $h \to b \bar{b}$  and $h\to gg$ partial widths (which will be discussed below), and therefore to the total width. These corrections alter the 
diphoton branching fraction, and therefore the signal strength, for a given value of $r_{\gamma\gamma}$. Note also that the distributions of 
$r_{\gamma\gamma}$ in the neutralino and gravitino model sets are rather similar yet somewhat distinct from the corresponding distribution in the low-FT model 
set, which exhibits a broader range of values for $r_{\gamma\gamma}$ despite the lower statistics. This larger spread in the low-FT distribution 
arises from the mandatory presence of light charginos, stops, and (in many cases) sbottoms, typically resulting in larger SUSY corrections to the effective 
$h \gamma\gamma$ coupling than in the large neutralino and gravitino model sets, in which charged sparticles are not required to be relatively light. Finally, 
note that in all three model sets the value of $r_{\gamma\gamma}$ peaks at the roughly same value, slightly above unity. We will see below that this shift 
is reasonably anticorrelated with a corresponding shift in the peak of the $r_{gg}$ distribution (as well as with the $ht\bar{t}$ coupling in the low-FT set).
Both offsets generally result from the large stop mixing that is necessary to obtain the correct value of the Higgs mass.

Figure~\ref{figC} displays analogous histograms of the ratio $r_{gg}$, showing the distribution for each pMSSM model set. Once again, we see that 
the neutralino and gravitino distributions are quite similar while the low-FT distribution differs as a result of distinct requirements on the sparticle 
spectra. Note that all three distributions peak below unity.
As shown in, \eg, \cite{Carena:2013iba}, the large Higgs mass generally requires large stop mixing in Supersymmetry, 
which results in a small ($\sim 5\%$) but important reduction in the $h \to gg$ partial width and a simultaneous, but somewhat smaller, enhancement in the $h \to \gamma\gamma$ partial width.  This is a consequence of the non-decoupling nature of SUSY corrections to the Higgs sector. If the stop 
sector radiative corrections were totally responsible for this deviation (which is a reasonable approximation in many cases), then the shift in $r_{gg}$ at the amplitude level would be 
$\sim 3$ times larger than the corresponding change in $r_{\gamma\gamma}$, with the two displacements having opposite signs. As a result of this effect, essentially all 
of our models predict $r_{gg}$ to be below unity; this observation will figure prominently in our subsequent discussion of future experimental constraints 
on the Higgs couplings. Interestingly, we also see that the tails of the $r_{gg}$ distribution are not very large for the neutralino and gravitino pMSSM model sets. 
The tails are slightly smaller in 
the low-FT $r_{gg}$ distribution, since the relevant corrections tend to be larger as a result of the bias towards light stops. Since 
the stops are playing an important role, we would expect corresponding shifts in the magnitude of the $ht\bar{t}$ coupling; as we will see below, 
this is indeed the case.

Figure~\ref{figD} displays the results for the ratio $r_{bb}$ for the three pMSSM model sets, with the neutralino and gravitino distributions again being similar, yet
somewhat different from the low-FT scenario. Small differences between the neutralino and gravitino distributions arise from several reasons, but namely from the fact that 
lighter stops/sbottoms can appear in the gravitino set, since the requirement for the stop to be heavier than the LSP is trivially satisfied when $m_{LSP} \sim 0$ 
as in most of the gravitino LSP models. For each pMSSM model set we see the now-familiar pattern in which the LHC direct SUSY searches do not significantly alter the shapes 
of the partial width distributions. Unlike the previous cases, however, we now see that $r_{bb}$ may deviate from unity by a significant $O(1)$ factor. These 
deviations result from large sbottom mixing that can make corresponding $O(1)$ changes in the $hb\bar b$ couplings through non-decoupling (mostly gluino) 
loop effects. These loop effects are driven by the size of the off-diagonal element of the sbottom mass matrix, \ie,  $m_b(A_b-\mu \tan \beta)$, which is 
enhanced for large values of $\tan \beta$. While the tails of this distribution mostly extend to larger values of $r_{bb}$, we see that models also exist 
with $r_{bb}$ being significantly below unity. Since the $b\bar b$ mode dominates the Higgs width, the large variations in $r_{bb}$ also explain the large spread in the  
distribution of $r_{total}$, presented in Fig.~\ref{figA}. In our neutralino and gravitino parameter scans, 
$|A_b|$ and $|\mu|$ are typically of a similar size while $\tan \beta$ has typical values that are $\mathcal{O}(10)$, so the $\mu \tan \beta$ term 
in the off-diagonal element dominates. However, in the low-FT set this is no longer applicable since the allowed size of $|\mu|$ (and therefore the sbottom mixing) is significantly reduced by naturalness requirements. Thus in the low-FT scenario we expect a considerably smaller range of values for $r_{bb}$, which agrees with the distributions shown in Fig.~\ref{figD}.

Figure~\ref{figE} displays the analogous results for the ratio $r_{\tau\tau}$ for the three different pMSSM model sets. Here we again see that the shapes of the 
$r_{\tau\tau}$ histograms are not significantly altered by the ATLAS direct SUSY searches at this level of statistics. We also see that the peak occurs at a ratio 
value that is slightly greater than unity (by $\sim 2\%$) with a significant tail extending to larger values. This is not surprising since there are also 
non-decoupling effects in the corrections to the $h\tau \tau$ vertex.  However, these corrections occur via electroweakino loops and are proportional to the $\tau$ mass. 
This implies that the effect of these non-decoupling terms should be relatively small when compared with their corresponding effect in the ratio $r_{bb}$, 
and that is indeed what we observe. Again, since this non-decoupling occurs via the off-diagonal $m_\tau (A_\tau -\mu \tan \beta)$ term in the stau mass 
matrix, these effects should be somewhat suppressed in the low-FT model set in comparison to the other pMSSM model sets, and this is demonstrated in 
Figure~\ref{figE}.

\begin{figure}[htbp]
\centerline{\includegraphics[width=4.5in]{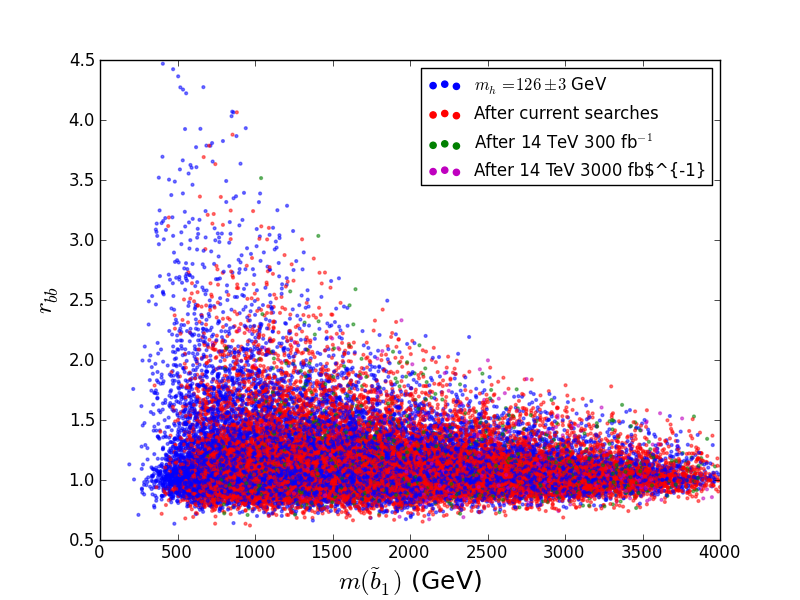}}
\vspace{-0.10cm}
\centerline{\includegraphics[width=4.5in]{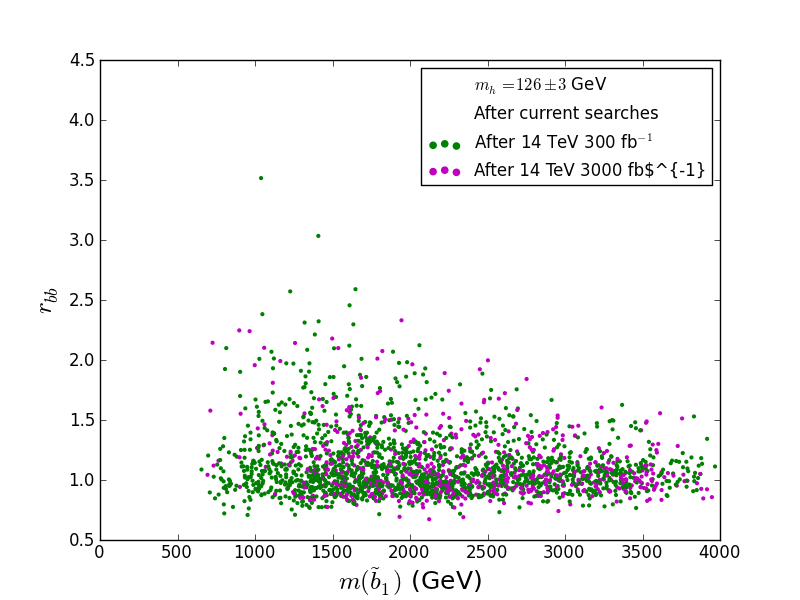}}
\vspace*{-0.10cm}
\caption{Values of the ratio $r_{bb}$ as a function of the lightest sbottom mass for the neutralino model set incorporating the influence of the ATLAS direct SUSY searches. 
The lower panel shows those models probed by the searches at 14 TeV. }
\label{figF}
\end{figure}
\begin{figure}[htbp]
\centerline{\includegraphics[width=4.5in]{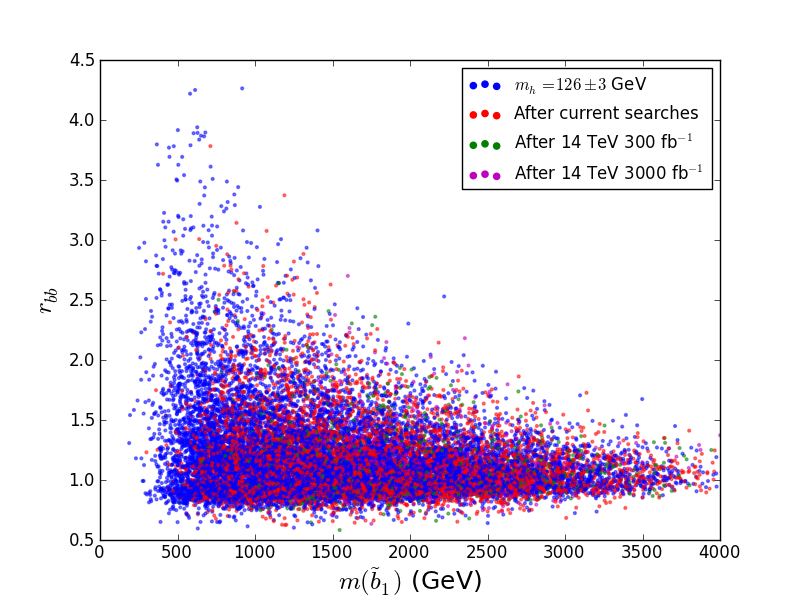}}
\vspace{-0.10cm}
\centerline{\includegraphics[width=4.5in]{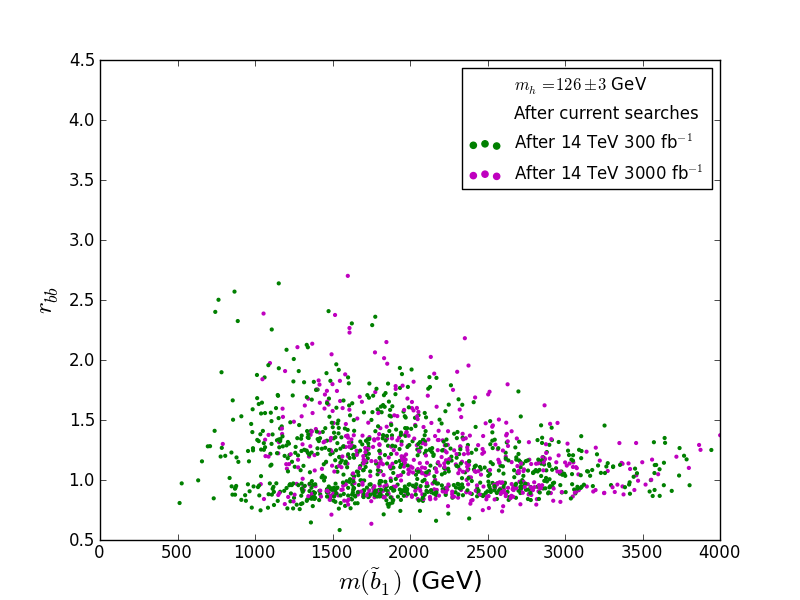}}
\vspace*{-0.10cm}
\caption{Same as the previous figure but now for the gravitino model set. }
\label{figF2}
\end{figure}

Figure~\ref{figF} shows the dependence of the ratio $r_{bb}$ on the lighter sbottom mass for the neutralino LSP model set with the effects of the direct LHC SUSY 
searches being imposed. Interestingly, measuring a value of this ratio near unity will not impose a constraint on the sbottom mass, regardless of the precision 
of the measurement. On the other hand, very large deviations of this ratio from unity are seen to require a relatively light sbottom mass, meaning that 
null SUSY search results should be able to reduce the expected range for $r_{bb}$. However, the non-decoupling nature of the corrections means that values 
of $r_{bb}$ above 2 are predicted, even after the 14 TeV direct SUSY searches are included.  Excluding O(1) deviations from $r_{bb} = 1$ (which can occur 
for sbottoms as heavy as 2.5 TeV) through direct SUSY searches is clearly not feasible. The large sbottom mass direct search reach necessary to constrain 
$r_{bb}$ significantly explains our earlier observation that this distribution is roughly independent of results from the LHC direct searches. 
Figure~\ref{figF2} shows that the corresponding results for the gravitino model sample are qualitatively similar, although they differ in detail due to the improved reach of direct sparticle searches 
in the gravitino set. Figure~\ref{figFP} shows the analogous results for the low-FT model set. As discussed above, the decreased range of $r_{bb}$ values in the low-FT model set arises from the requirement that $|\mu|$ is relatively small, decreasing the size of the off-diagonal 
element in the sbottom mass matrix and therefore the corrections to $r_{bb}$.

\begin{figure}[htbp]
\centerline{\includegraphics[width=4.5in]{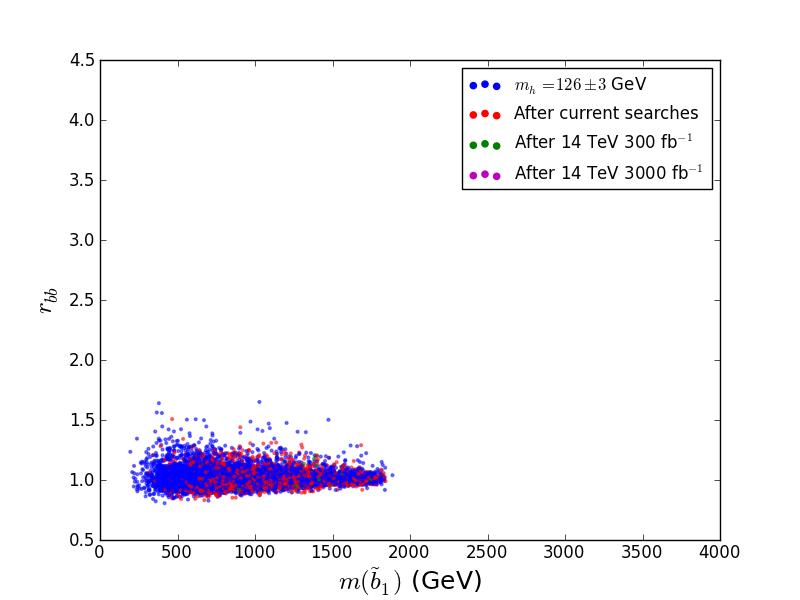}}
\vspace{-0.10cm}
\centerline{\includegraphics[width=4.5in]{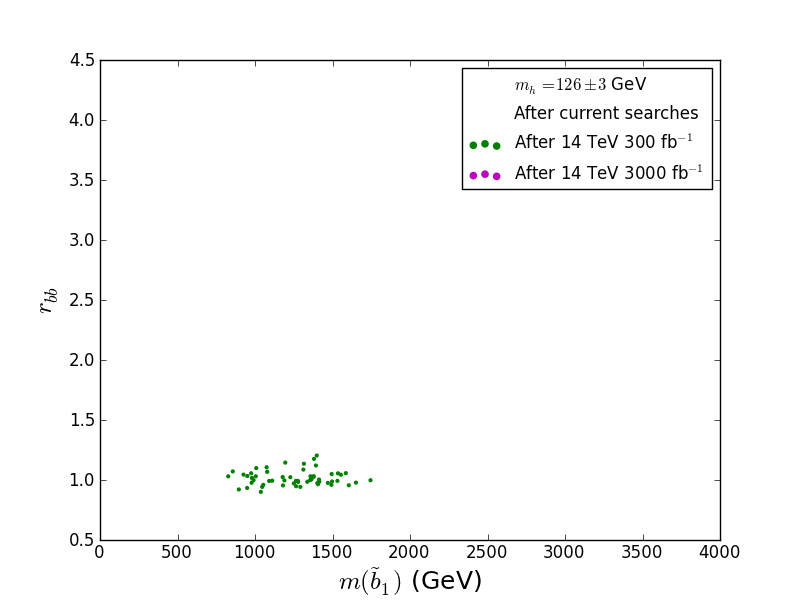}}
\vspace*{-0.10cm}
\caption{Same as the previous Figure but now for the low-FT model set.}
\label{figFP}
\end{figure}

If the lightest neutralino is sufficiently light, then the Higgs can decay to neutralino pairs, being observed as an invisible decay mode of the Higgs. 
The top panel in Fig.~\ref{figG} displays the branching fraction,
$B(h\to \chi\chi)$, as a function of the LSP mass for the few 
neutralino LSP models where this channel 
is kinematically allowed, and also indicates the influence of the direct SUSY searches at the LHC. Note that all of these models have values 
of $B(h\to \chi\chi) <0.5$, meaning that they remain allowed by the current LHC constraints on invisible Higgs decays. Since these models are mostly bino-Higgsino 
admixtures (to satisfy the WMAP/Planck relic density upper bound) and the coupling to the Higgs is proportional to the product of the bino and Higgsino content of the 
neutralino, the branching fractions are seen to fall rapidly as the neutralino mass increases. This is due not only to a reduction in phase space, but also to a 
decline in the neutralino Higgsino content as its mass increases. We note that all of these models will eventually be excluded (or discovered) 
by sparticle searches, as well as by searches for Higgs $\to$ invisible at the 14 TeV LHC and/or ILC500. The lower left panel shows the corresponding results 
for the gravitino set with a neutralino NLSP; here we see that a much smaller branching fraction is obtained since the WMAP/Planck constraint 
does not apply to the neutralino NLSP. Of course for these gravitino pMSSM models the lightest neutralinos will only produce an invisible final state if they escape the 
detector before decaying.  Neutralinos with $c \tau \lesssim 1$ m will have visible decays, generally producing a (possibly displaced) diphoton + MET signature, 
where the diphotons would of course fail to reconstruct the Higgs mass. However, the stability of the neutralino tends to be unimportant, since (with the possible 
exception of the models with very light neutralinos) the $h\to \chi \chi$ branching fraction is far too small to be accessible at the 14 TeV LHC. 
The bottom right panel displays the same distribution for the low-FT model set. Here we see that the additional constraints imposed on the pMSSM spectrum 
during the model generation yield numerous light LSPs that are mainly bino-Higgsino admixtures, a sizable fraction of which pair-annihilate via the Z/Higgs funnel. Note that 
these fall into two distinct branches, depending on the sign of the parameter $\mu$. In all cases, however, the invisible branching fraction is found to be 
below $\sim 30-50\%$, which will eventually be accessible at the 14 TeV LHC. While many of these models are now excluded by LHC direct SUSY searches, the remainder would 
be probed by the corresponding 14 TeV direct searches.

\begin{figure}[htbp]
\centerline{\includegraphics[width=5.5in]{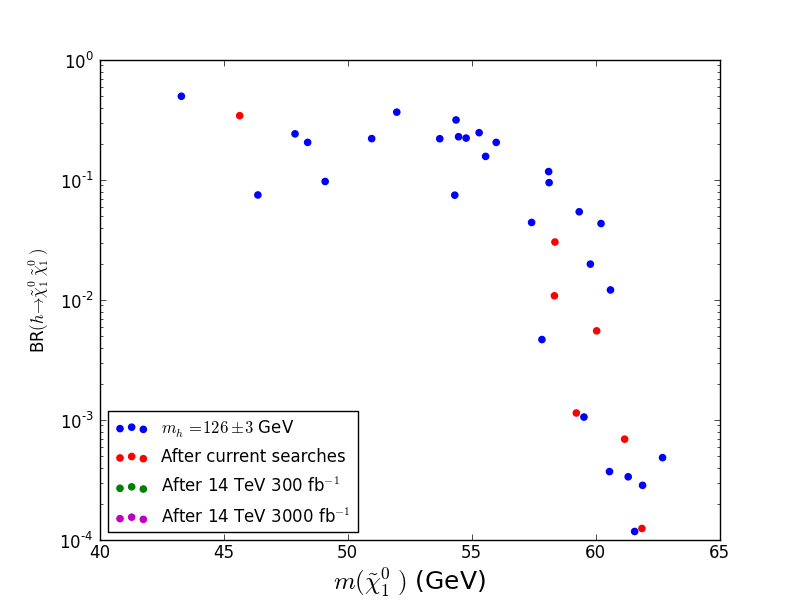}}
\vspace*{0.50cm}
\centerline{\includegraphics[width=3.5in]{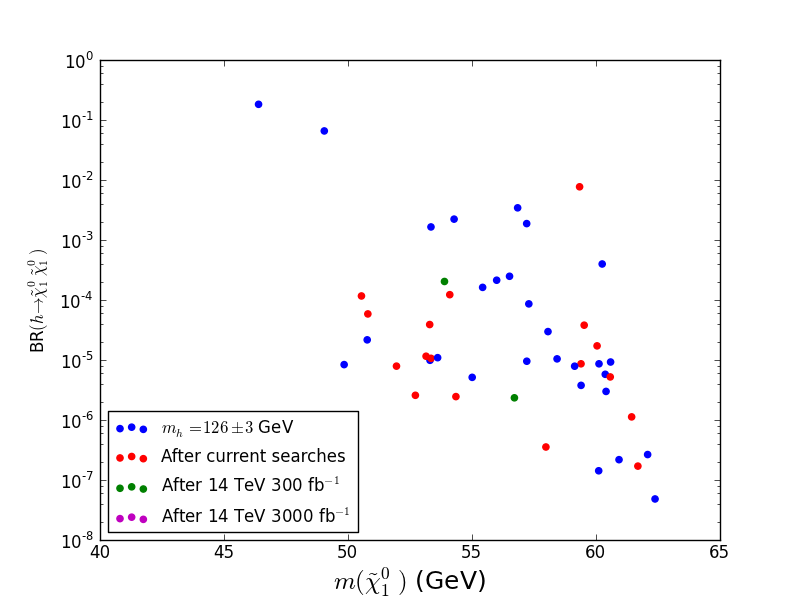}
\hspace{-0.50cm}
\includegraphics[width=3.5in]{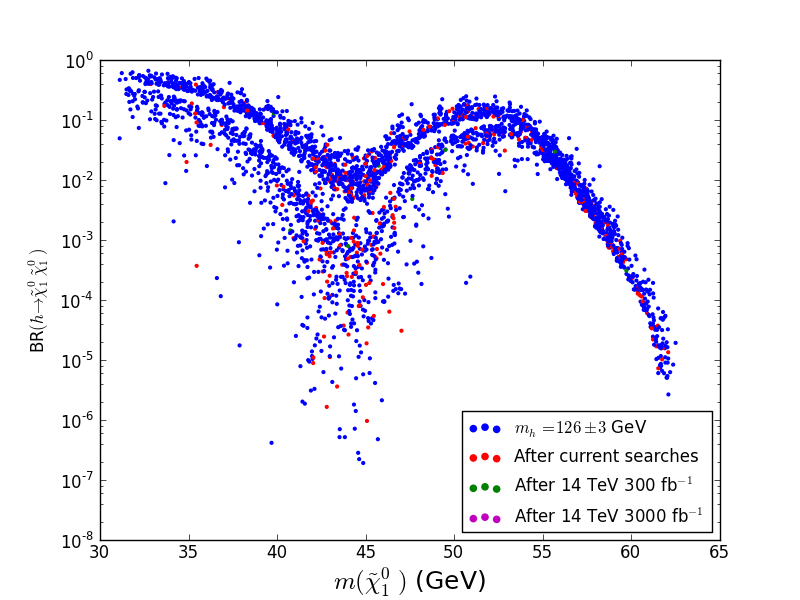}}
\vspace*{-0.10cm}
\caption{Branching fraction of the invisible Higgs decay $h \to \tilde{\chi}_1^0 \tilde{\chi_1}^0$ for neutralino LSP models (top) with the observed Higgs mass. The points 
are color-coded according to their coverage by the LHC direct SUSY searches. The analogous results for the gravitino (bottom left) in the case of a neutralino 
NLSP, and for the low-FT (bottom right) model sets are also shown.}
\label{figG}
\end{figure}

As a final observable, we briefly consider the ratio $r_{tt}$, defined as the squared value of the $ht\bar{t}$ coupling normalized to its SM value. We calculate this quantity using the expressions given in 
Ref.~\cite{Draper:2013oza}. The predicted values of this ratio, computed for each of the various model sets before the application of constraints from the 
SUSY direct searches, are of some interest for future measurements at both the 14 TeV LHC and at the ILC.  They are displayed in Figure~\ref{figH}. 
Here we see that the deviation from the SM expectation is always less than $\sim 10\%$, which is below the anticipated sensitivity of both LHC14 
and ILC500. However, the 1 TeV upgrade of the ILC should eventually be able to determine this quantity at the level of a few percent{\cite {ILC}}. 
The different behavior of the histograms for the three model sets is easily understood when we recall that the deviations from unity are driven mostly by the 
non-decoupling effects of the stop masses and, particularly, by the mixing in the stop sector that are controlled by the values of the parameters $A_t, \mu$ and 
$\tan \beta$ via the quantity $X_t \sim A_t-\mu /\tan \beta$.  If there 
is no strong preference for the sign of $A_t$ and/or $\mu$ arising from the model generation procedure this distribution will be approximately 
symmetric around the SM value; this is observed for the neutralino set. We note that while $A_t$ is, in fact, sign symmetric, the corresponding 
distribution of the values of $\mu$ is found to be somewhat asymmetric in sign when $|\mu|$ is large. However, most of the models in this set have 
relatively small values of $|\mu|$ (as Higgsino LSPs are common) so that the resulting distribution remains essentially symmetric, as shown in the Figure.
For the gravitino set, these same conditions hold except that Higgsino LSPs are somewhat less common and the values of $|\mu|$ tend to be correspondingly larger,
thus $r_{tt}$ is now more sensitive to this sign asymmetry in the $\mu$ distribution. Hence, for the gravitino set, we see a somewhat asymmetric distribution for $r_{tt}$. 
The low-FT model set displays a different behavior, as here $|\mu|$ must be small and we simultaneously require both the observed 
value of the Higgs mass and also less than than $1\%$ values of fine-tuning. This selects a specific sign for the stop mixing as well as a hierarchical 
stop spectrum. This pushes $r_{tt}$ to somewhat larger deviations from the SM, on average, than in the other two model sets, with a strong preference towards increasing the $ht\bar{t}$ coupling with respect to its Standard Model value.

\begin{figure}[htbp]
\centerline{\includegraphics[width=6.5in]{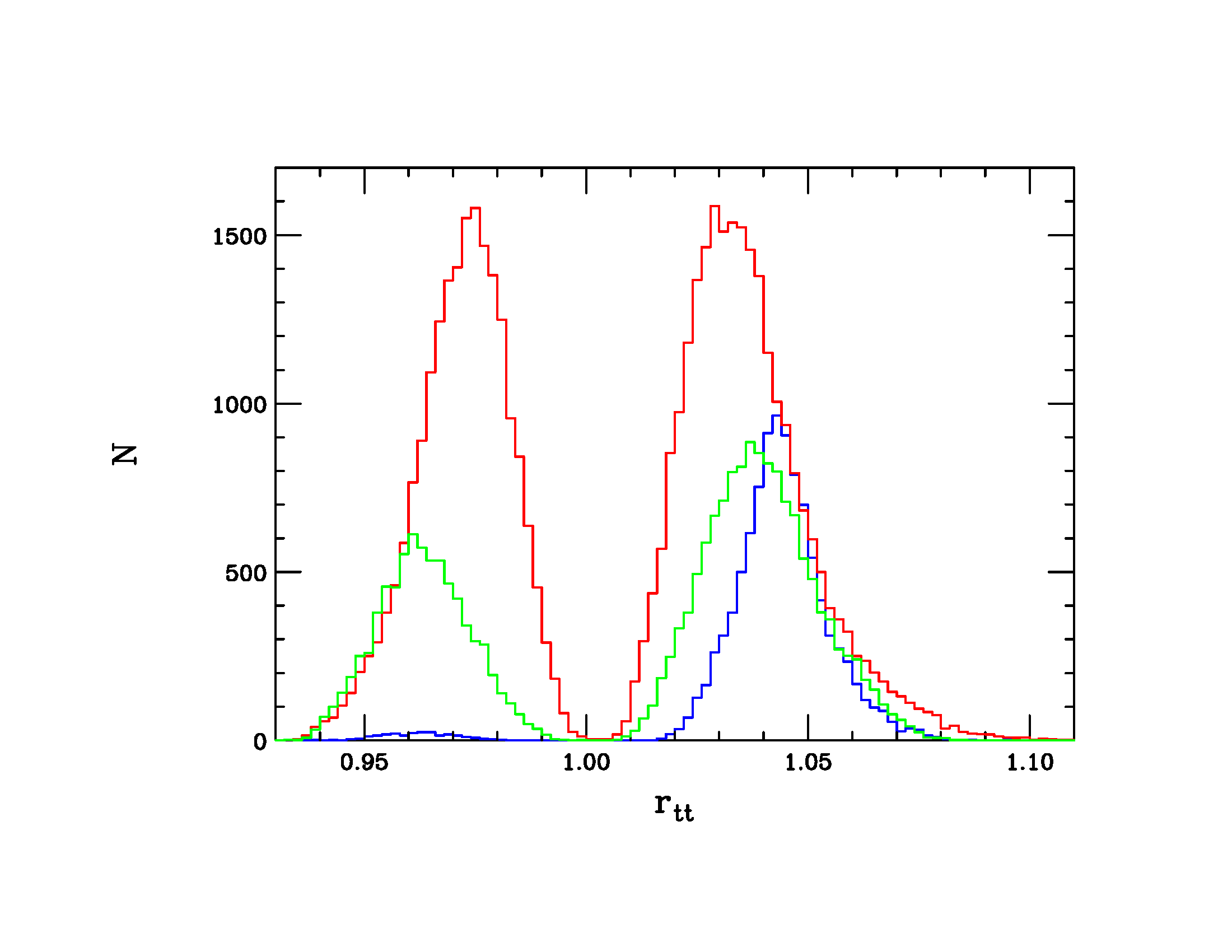}}
\vspace*{-0.10cm}
\caption{Histograms of the predicted values of the ratio $r_{tt}$, defined in the text, for the various pMSSM model 
sets: neutralino (red), gravitino (green) and low-FT (blue).}
\label{figH}
\end{figure}

\section{Analysis and Results} 
\label{sec:step3}

Now that we have assembled the necessary ingredients, we can determine how the future measurements of the various Higgs couplings at the LHC and ILC will 
restrict the pMSSM parameter space, and compare these constraints with those from the direct SUSY searches at the LHC.  In this analysis, we use the 
numerical results for the current and expected future Higgs coupling measurements at the LHC and ILC 
presented in Refs.~\cite{now,14TeVLHC,ILC}.  We note that important higher-order corrections to the Higgs couplings have yet to be computed, and these
may be quite relevant compared to the claimed level of precision for the future collider measurements.  We are thus unfortunately forced to ignore these potentially
significant theoretical uncertainties in quoting allowed ranges for the ratios of Higgs couplings in the pMSSM to those in the SM.
We remind the reader to keep this important issue in mind when interpreting our results, and note they should be treated as indicative only. Clearly, 
more theoretical work will be necessary before (sub-)percent-level measurements are truly meaningful. 

We also caution the reader that in obtaining the results shown below, we have necessarily made an assumption about the central value of the 
future Higgs coupling measurements at the LHC and ILC.
Namely, we have assumed that the central values will coincide {\it exactly} with those predicted by the SM, {\it i.e.}, we take the measured central value to be $r_{X}=1$
for all couplings. As we will see from the discussion that follows,  the observation of Higgs couplings not centered around the SM prediction
(even within the expected ranges) would probe a different fraction of our 
model sets. This is particularly true for the case of the couplings generated at loop-level, $r_{gg}$ and $r_{\gamma\gamma}$, where the pMSSM predictions deviate from the SM values 
essentially all in one direction. Of course, our qualitative results, which indicate that precise Higgs coupling measurements (when properly 
understood) have significant sensitivity to the pMSSM, do not depend on the actual central values that will be observed for these couplings.  

Assuming that the future measured central value of each parameter is equal to its SM prediction, we return to Figs.~\ref{figB},~\ref{figC},~\ref{figD} and~\ref{figE} 
(as well as~\ref{figG}) and now concentrate on the vertical lines, which display the expected sensitivity arising from  future experiments. These show the regions of the 
various $r_X$ that will be allowed or excluded at the $95\%$ CL by Higgs coupling measurements at the LHC and HL-LHC{\cite {14TeVLHC}}, 
the 500 GeV ILC (ILC500) and the ILC500 with a luminosity upgrade{\cite {ILC}}, here denoted as HL-ILC500.  Of course, it is important to once again note that  
these future expected allowed regions can always be shifted, allowing for the estimation of implications of other possible experimental outcomes. 
We first notice that current LHC data on the Higgs couplings does not significantly constrain the pMSSM parameter space, since the precision of the Higgs measurements 
is still rather low in comparison to the deviations expected in the pMSSM. Once 14 TeV LHC, as well as ILC, data is available, this will no longer 
be the case and the measurements will begin to probe pMSSM effects as their accuracy improves.
However, the key result here is that, regardless 
of what central values are actually observed, {\it indirect Higgs coupling measurements will likely result in the exclusion (or the discovery) of pMSSM models 
that are not accessible to the direct SUSY searches at the LHC}. An important caveat to this, of course, is that we need to include the full suite of 14 TeV direct SUSY 
searches before this result is truly robust. However, given our 7 and 8 TeV studies~\cite{usnew}, the 0-$\ell$ jets plus MET search when combined 
with the 0,1-$\ell$ stop searches will result in powerful parameter space coverage at 14 TeV, and so this qualitative conclusion is unlikely to change. This result 
is found to hold for all of the model sets.

Taking these results at face value, we can extract some relevant numbers directly from these Figures. We can now determine
what fraction of the presently allowed pMSSM models, \ie, those passing the 7 and 8 TeV ATLAS direct search analyses (with $m_h=126\,\pm\,3$ GeV), 
will be indirectly probed by future measurements of the Higgs couplings.  Next, we can ascertain how 
these results will be modified by the 14 TeV LHC direct SUSY searches. Our results are presented in the set of 
Tables~\ref{T1},~\ref{T2} and~\ref{T3} for both the 14 TeV LHC and the ILC. 
In these Tables we see a number of important results: ($i$) at the LHC, constraining the $hb\bar b$ coupling yields the 
strongest bounds on the allowed pMSSM parameter space. This measurement can be greatly improved at the ILC, which has the potential to yield exquisite precision on this coupling.
($ii$) However, given our assumption that the measured central values exactly correspond to the SM predictions, we see that the ILC determination of the $hgg$ coupling probes much,
if not all, of the remaining pMSSM parameter space. The reason for this is clear: Since $r_{gg}$ is forced to be less than unity by the non-decoupling effects associated with
the large stop mixing required 
to generate the observed value of the Higgs mass, a determination of $r_{gg}=1$ with a very small error will probe essentially all of the model sets! If, on the other 
hand, the central value were measured to be, say, only $\sim 2-3\%$ below unity, a very much smaller fraction of models would then be probed. For example, 
if the central value of $r_{gg}$ were measured to be 0.97 with the same expected errors, then we find that this measurement is only sensitive to $2.7\%$ of the 
neutralino LSP model set at the ILC500, so that $hb\bar b$ would remain the dominant constraint in this case. This 
specific example demonstrates the sensitivity of our results to the assumption that the measured central values will always agree with the SM predictions. In any case, 
($iii$) we see that both the LHC and ILC will provide very powerful probes of the pMSSM model space and have the potential to observe the effects of at least some of 
the models that would otherwise remain viable, being missed by the 14 TeV direct SUSY searches. In particular, the precision attainable in Higgs coupling measurements at the ILC 
will deeply probe the pMSSM parameter space.

Tables~\ref{T1}-\ref{T3} also show that ($iv$) although the general shapes of the $r_X$ distributions are somewhat similar, they differ in detail so  
that the three pMSSM model sets will respond distinctly to constraints from the various indirect Higgs coupling measurements. Of course, the ILC500 
is extremely powerful in all three cases.  The last thing we notice is that ($v$) the entries in the Tables will not vary greatly as we include more channels 
from future direct SUSY searches at the 14 TeV LHC. This is not surprising; in the limit that the shapes of the $r_X$ distributions are completely unaffected by the 
SUSY search results, the Table entries should be essentially independent of which LHC searches have been applied. The limited statistical size of our model samples, and 
the small changes in the $r_X$ distribution shapes, account for the observed variations.

\begin{table}
\begin{center}
\begin{tabular}{|c||c|c|c|c|}
\hline
Channel & 300 fb$^{-1}$ LHC & 3 ab$^{-1}$ LHC & 500 GeV ILC & HL 500 GeV ILC \\
\hline
\hline

$b\bar{b}$ & 16.6 (27.7, 0.5) & 33.4 (48.5, 5.5) & 78.4 (88.8, 49.1) & 91.1 (95.8, 77.3) \\
$\tau \tau$ & 0.7 (0.8, 2.9) & 3.1 (2.7, 5.7) & 11.5 (9.9, 11.9) & 36.9 (34.2, 32.9) \\
$gg$ & 0.02 (0.04, 0.5) & 0.5 (0.6, 3.1) & 99.4 (99.7, 99.7)& 100.0 (100.0, 100.0) \\
$\gamma \gamma$ & 0.02 (0.07, 0) & 0.02 (0.09, 0.2) & 0.02 (0.07, 0) & 0.1 (0.2, 0.6) \\

Invisible & 0 (0, 0) & 0 (0, 0) & 0.01 (0.01, 6.2) & 0.02 (0.01, 7.5) \\
\hline
\hline

All & 17.1 (28.2, 3.8) & 34.9 (49.6, 11.1) & 99.8 (99.96, 99.92) & 100.0 (100.0, 100.0) \\
\hline
\end{tabular}
\caption{The fraction in percent of the neutralino (gravitino, low-FT) model sets (with the correct Higgs mass), which remain viable after the current 7 and 8 TeV LHC searches, 
that can be probed by future Higgs coupling measurements, {\it assuming} that the SM values for these couplings are observed.}
\label{T1}
\end{center}
\end{table}
\begin{table}
\begin{center}
\begin{tabular}{|c||c|c|c|c|}
\hline
Channel & 300 fb$^{-1}$ LHC & 3 ab$^{-1}$ LHC & 500 GeV ILC & HL 500 GeV ILC \\
\hline
\hline

$b\bar{b}$ & 20.5 (31.7, 0) & 39.1 (53.0, 5.4) & 82.6 (92.6, 46.4) & 93.1 (97.5, 75.0) \\
$\tau \tau$ & 0.5 (0.7, 1.8) & 3.3 (2.3, 1.8) & 12.9 (9.9, 5.4) & 38.9 (32.6, 23.2) \\
$gg$ & 0 (0, 0) & 0.09 (0.1, 0) & 99.9 (99.93, 100.0) & 100.0 (100.0, 100.0) \\

$\gamma \gamma$ & 0 (0, 0) & 0 (0, 0) & 0 (0, 0) & 0 (0, 0) \\

Invisible & 0 (0, 0) & 0 (0, 0) & 0 (0, 10.7) & 0 (0, 16.1) \\
\hline
\hline

All & 20.8 (31.9, 1.8) & 40.6 (53.7, 5.4) & 99.91 (100.0, 100.0) & 100.0 (100.0, 100.0) \\
\hline
\end{tabular}
\caption{Same as Table~\ref{T1} above, but now for the subset of models expected to remain viable after the ATLAS 14 TeV 0l jets + MET and 0l and 1l stop searches with 300 fb$^{-1}$ of 
integrated luminosity.}
\label{T2}
\end{center}
\end{table}
\begin{table}
\begin{center}
\begin{tabular}{|c||c|c|c|c|}
\hline
Channel & 300 fb$^{-1}$ LHC & 3 ab$^{-1}$ LHC & 500 GeV ILC & HL 500 GeV ILC \\
\hline
\hline

$b\bar{b}$ & 19.6 (32.6, ---) & 38.4 (54.5, ---) & 82.9 (94.9, ---) & 93.4 (98.4, ---) \\
$\tau \tau$ & 0.7 (0.7, ---) & 3.3 (2.5, ---) & 14.7 (10.7, ---) & 41.6 (35.3, ---) \\
$gg$ & 0 (0,---) & 0 (0, ---) & 100.0 (100.0, ---) & 100.0 (100.0, ---) \\

$\gamma \gamma$ & 0 (0, ---) & 0 (0, ---) & 0 (0, ---) & 0 (0, ---) \\

Invisible & 0 (0, ---) & 0 (0, ---) &  0 (0, ---) & 0 (0, ---) \\
\hline
\hline

All & 29.9 (32.8, ---)  & 39.3 (55.4, ---) & 100.0 (100.0, ---) & 100.0 (100.0, ---) \\
\hline
\end{tabular}
\caption{Same as Table~\ref{T1} above but now for the subset of models expected to remain viable after the ATLAS 0l jets + MET and 0l and 1l stop searches with 3 ab$^{-1}$ of 
integrated luminosity. The entries for the low-FT set in this table are blank because no models survive the 3 ab$^{-1}$ LHC direct searches.}
\label{T3}
\end{center}
\end{table}

Lastly, we summarize our results in the $M_A - \tan\beta$ plane by combining the effects of anticipated Higgs coupling measurements in the $\gamma\gamma\,, \tau\tau\,, b\bar b$ channels
for the neutralino LSP model set.  
We exclude the $hgg$ coupling from this analysis due to the complications and resulting shift in the central value of this parameter arising from the large stop mixing as discussed above.
Figure ~\ref{newplot}
shows the fraction of models in the large neutralino LSP sample that are probed in a particular bin by the anticipated measurements of these three channels
at the LHC and ILC500 and their luminosity upgrades.  The fraction is color-coded, indicating the pMSSM coverage within a bin, ranging from 100\% (black)
coverage to 0\% (dark blue).  The white curves represent the results from current heavy Higgs searches with decays into $\tau$ pairs~\cite{CMSextra}.  Here, we clearly see the
effects of increasing precision for the Higgs coupling measurements, and the value of the anticipated ultra-precise determinations to be available at the ILC500, in covering the pMSSM
parameter space.  We note that the Higgs coupling measurements cover a region of parameter space that is somewhat orthogonal to that of the heavy Higgs searches.  Namely,
the coupling determinations probe essentially vertical slices of this plane, and most importantly, catch the low $M_A$, $\tan\beta$ region that is missed by the direct searches.  This
demonstrates the complementarity of the direct and indirect approaches.

\begin{figure}[htbp]
\centerline{\includegraphics[width=3.5in]{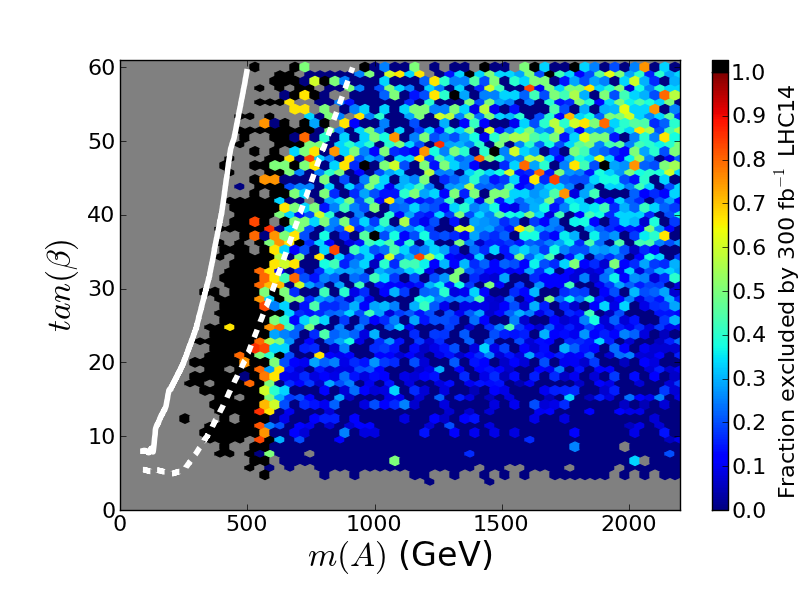}
\hspace{-0.50cm}
\includegraphics[width=3.5in]{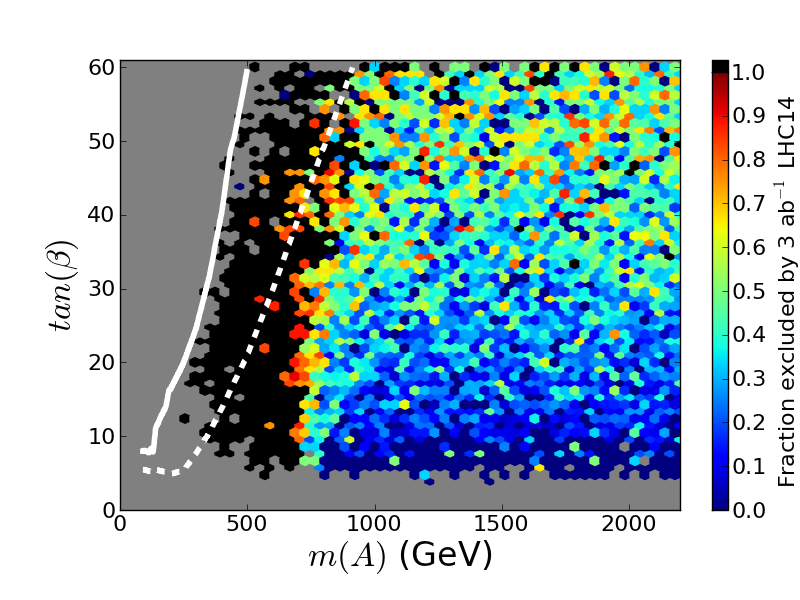}}
\vspace*{0.50cm}
\centerline{\includegraphics[width=3.5in]{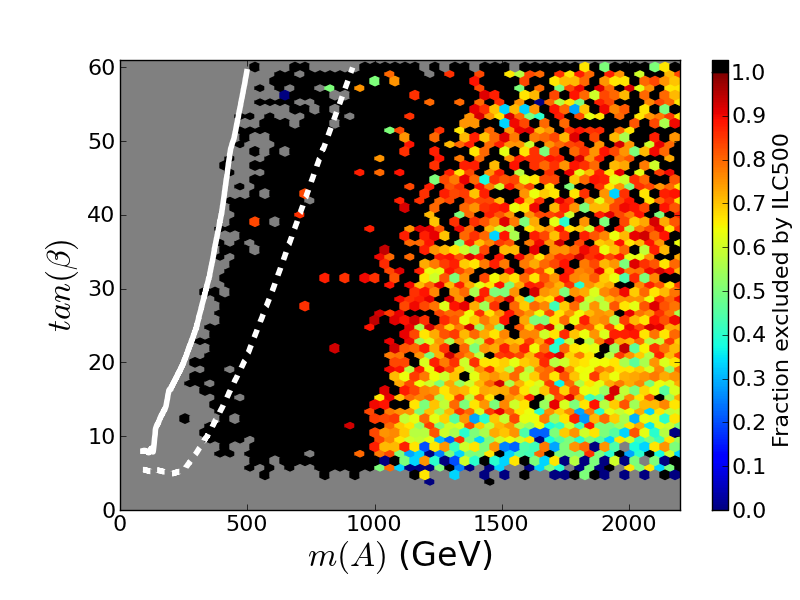}
\hspace{-0.50cm}
\includegraphics[width=3.5in]{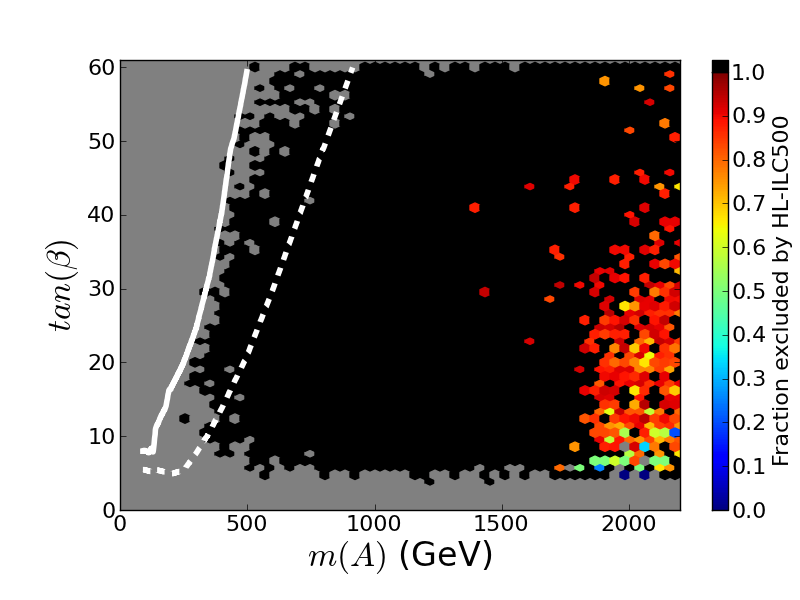}}
\vspace*{-0.10cm}
\caption{Coverage of the pMSSM parameter space for the neutralino model set in the $M_A - \tan\beta$ plane, 
showing the fraction of models probed in each bin by the anticipated
sensitivity to the combined $\gamma\gamma$, $\tau\tau$, and $b\bar b$ Higgs couplings at various colliders as indicated.  The white curves correspond to the
present limits from the direct searches for $H/A\to\tau\tau$.}
\label{newplot}
\end{figure}

\section{Conclusion}

In this paper we have examined SUSY signals and Higgs boson properties within the context of the pMSSM for models with either neutralino or gravitino LSPs as 
well as for neutralino models with low FT that saturate the WMAP/Planck relic density measurement. Within this general scenario we then addressed the following 
questions: `What will potentially null searches for SUSY at the LHC tell us about the possible properties of the Higgs boson?' and, conversely, `What do the 
precision measurements of the couplings of the Higgs tell us about the possible properties of the various superpartners?' We again note that in 
obtaining the results presented here we have ignored any theoretical errors associated with the as-yet  to be computed corrections for the Higgs coupling ratios,
and employed the current version of the corrections as implemented in HDECAY. 
Our results can be further refined once a better understanding of this uncertainty is provided by future theoretical work. 

We saw in the above discussion that the answer to the first question was rather straightforward: Given an initial distribution of signal strengths $\mu_X$ or 
branching fraction ratios $r_X$ for a specific final state, the LHC direct SUSY searches reduce the size of the distribution but to a very good approximation 
do not change its {\it shape}. This was shown to be true for all three of the model sets we consider. This implies that to first order the direct (null) SUSY 
searches at the LHC will not impact the range of possible deviations of Higgs branching fractions from their SM values. This is a very powerful result.  

However, we found the answer to the second question to be much more complex and of potentially of even greater importance: Precision measurements of Higgs 
couplings and branching fractions can and do lead to the exclusion of pMSSM models which cannot be probed by the powerful 14 TeV LHC direct SUSY searches, even 
with an integrated luminosity of 3 ab$^{-1}$. This is true for both gravitino and neutralino model sets and also true whether or not the precise values of 
the measured quantities are consistent with the SM expectation. Of course, the more precisely the Higgs couplings are measured, the greater the fraction of 
the pMSSM models that can be probed. Since the $hb\bar b$ coupling can deviate the furthest from its SM value within the pMSSM framework, measurements of 
its value generally will have the greatest impact {\it if} we do not assume that the central values measured for the Higgs couplings are given exactly by 
their SM values. If this is indeed the case, however, then the $hgg$ coupling at the ILC will provide the strongest constraint as this quantity is necessarily 
shifted in the pMSSM by stop loops with a central value crudely determined by the requirement of obtaining the observed Higgs mass. In such a case (or if the 
observed central values for $r_{gg}$ -- or to a lesser extent $r_{\gamma\gamma}$ -- differ from the SM in the opposite direction from the pMSSM prediction), 
essentially all of the pMSSM parameter space considered here would then be excluded.  

Lastly,
we compared the reach of the Higgs coupling determinations to the direct heavy Higgs searches in the $M_A - \tan\beta$ plane
and show that they cover orthogonal regions.  

Our analysis
demonstrates the complementarity of the direct and indirect approaches in searching for Supersymmetry, and the importance of precision studies of the properties of the
Higgs Boson.

\section{Acknowledgments}

We wish to thank M.~Spira for answering our many questions and for assistance with the implementation of HDECAY. This work was 
supported by the Department of Energy, Contracts DE-AC02-06CH11357, DE-AC02-76SF00515, and DE-FG02-12ER41811.

\end{document}